\documentclass[twocolumn]{aastex631}

\usepackage{amsmath}	
\usepackage{amssymb}	
\usepackage{wasysym}
\usepackage{multirow}
\usepackage[flushleft]{threeparttable}
\usepackage{booktabs}

\newcommand{\MHupdate}[1]{{#1}} 
\shorttitle{Ensemble Mixed-mode Asteroseismology}
\shortauthors{Kuszlewicz et al.}

\graphicspath{{./}{figures/}}

\begin{document}

\title{Mixed-mode Ensemble Asteroseismology of Low-Luminosity \textit{Kepler} Red Giants}

\correspondingauthor{James S. Kuszlewicz}
\email{james.kuszlewicz@gmail.com}

\author[0000-0002-3322-5279]{James S. Kuszlewicz}
\affiliation{Center for Astronomy (Landessternwarte), Heidelberg University, K\"onigstuhl 12, 69118 Heidelberg, Germany\\}
\affiliation{Stellar Astrophysics Centre, Department of Physics and Astronomy, Aarhus University, Ny Munkegade 120, DK-8000 Aarhus C, Denmark\\}
\affiliation{Max-Planck-Institut f\"{u}r Sonnensystemforschung, Justus-von-Liebig-Weg 3, 37077 G\"{o}ttingen, Germany}

\author[0000-0003-2400-6960]{Marc Hon}
\altaffiliation{NASA Hubble Fellow}
\affiliation{Institute for Astronomy, University of Hawai`i,
2680 Woodlawn Drive, Honolulu, HI 96822, USA}

\author[0000-0001-8832-4488]{Daniel Huber}
\affiliation{Institute for Astronomy, University of Hawai`i,
2680 Woodlawn Drive, Honolulu, HI 96822, USA}

\begin{abstract}

We present measurements of the dipole mode asymptotic period spacing ($\Delta\Pi_1$), the coupling factor between p- and g- modes ($q$), the g-mode phase offset ($\epsilon_g$), and the mixed-mode frequency rotational splitting ($\delta\nu_{\mathrm{rot}}$) for 1,074 low-luminosity red giants from the \textit{Kepler} mission. Using oscillation mode frequencies extracted from each star, we apply Bayesian optimization to estimate $\Delta\Pi_1$ from the power spectrum of the stretched period spectrum and to perform the subsequent forward modelling of the mixed-mode frequencies.
With our measurements, we show that the mode coupling factor $q$ shows significant anti-correlation with both stellar mass and metallicity, and can reveal highly metal-poor stars.
We present the evolution of $\epsilon_g$ up the lower giant branch up to before the luminosity bump, and find no significant trends in $\epsilon_g$ or $\delta\nu_{\mathrm{rot}}$ with stellar mass and metallicity in our sample. Additionally, we identify six new red giants showing anomalous distortions in their g-mode pattern. Our data products, code, and results are provided in a public repository.


\end{abstract}



\section{Introduction} \label{sec:intro}
With uninterrupted, high-precision photometry, the \textit{Kepler} space mission \citep{Borucki_2010} has brought vast insights into the interiors of red giant stars. Oscillation modes exhibiting properties comprising a hybrid of acoustic (p-) and gravity (g-) modes, which are known as mixed-modes, have offered diagnostics into the properties of stellar evolution that would have otherwise been impossible to measure. These include the differentiation between hydrogen shell-burning and core helium-burning giants (e.g., \citealt{Bedding_2011, Corsaro_2012, Stello_2013, Mosser_2014, Vrard_2016, Elsworth_2017, Hon_2017, Hon_2018, Elsworth_2019}), core rotation rates (e.g., \citealt{Beck_2012, Mosser_2012, Mosser_2015, Di_Mauro_2016, Triana_2017, Di_Mauro_2018, Gehan_2018, Tayar_2019, Deheuvels_2020}), stellar inclination angles (e.g., \citealt{Huber_2013, Campante_2016, Corsaro_2017, Mosser_2018, Kuszlewicz_2019, Gehan_2021}),
as well as structural variations within the deep stellar interior \citep{Cunha_2015, Mosser_2015, Vrard_2019, Vrard_2022}.

Low-luminosity ($L\apprle20L_{\odot}$) red giants are particularly useful targets for probing stellar interiors because they have non-negligible coupling strengths between their p- and g-mode cavities and have modest mixed-mode densities within the observable range of their oscillation frequencies. A convenient formulation for the mixed-modes of such stars is the asymptotic relation for p-g mixed-modes \citep{Shibahasi_1979, Unno_1989}. As described by \citet{Mosser_2012b}, the formulation describes the coupling of p-g modes using the dipole asymptotic g-mode period spacing ($\Delta\Pi_1$), the coupling factor between p- and g- modes ($q$), and a phase offset ($\epsilon_g$) describing the behaviour of modes near their interior turning points (e.g., \citealt{Hekker_2017}).

Parameterizing the mixed-mode pattern for red giants therefore requires a search over $\Delta\Pi_1$, $q$, and $\epsilon_g$. To model several red giants, 
\citet{Buysschaert_2016} constructed grids requiring millions of discrete points over the tridimensional ($\Delta\Pi_1$, $q$, $\epsilon_g$) parameter space. The study additionally identified that such a parameter space was strongly multi-modal, which makes it challenging to explore with conventional optimization algorithms.

Another complicating factor to dipole mixed-mode parameterization is the frequency splitting of mixed-modes due to effects from rotation. To address this, \citet{Mosser_2015} introduced stretched periods, which has successfully enabled the identification of rotationally-split mixed-modes \citep{Mosser_2015} and the measurement of core rotation rates \citep{Gehan_2018}. The inclusion of rotational splittings $\delta\nu_{\mathrm{rot}}$ as a fourth parameter expands the 
number of candidate points in the optimization task by an order of magnitude and adds to the search complexity due to the required coordinate transformation from frequency to to stretched periods.
A search over the quadridimensional ($\Delta\Pi_1$, $q$, $\epsilon_g$, $\delta\nu_{\mathrm{rot}}$) space is therefore a challenging task. To date, the total number of red giants for which these four mixed-mode parameters have
been estimated using forward models of dipole mixed modes is small $\sim200$; e.g., \citealt{Mosser_2018}). 

Similar to ensemble studies of red giant using p-mode asteroseismology (e.g., \citealt{Yu_2018}), we seek to facilitate ensemble studies of mixed-mode asteroseismology. Here, we extracted the mode frequencies of over 1,000 \textit{Kepler} red giants and estimate their mixed-mode properties using Bayesian optimization to search over the quadridimensional mixed-mode parameter space.


\section{Methods}
\subsection{Mixed-mode Formalism}
The p-g mode asymptotic expansion as expressed by \citet{Mosser_2012b} is as follows:

\begin{equation} \label{eqn:asymptotic_mixed_modes}
    \nu_m = \nu_{n_p} + \frac{\Delta\nu}{\pi} \arctan \bigg[q\tan\pi\bigg(\frac{1}{\Delta\Pi_1\cdot\nu_m}-\epsilon_g\bigg)\bigg],
\end{equation}
where $\nu_{n_p}$ is the frequency of the p-mode (with radial order $n_p$) to which the g-mode couples with, $\Delta\nu$ is the large frequency separation of p-modes, $\Delta\Pi_1$ is the dipole asymptotic g-mode period spacing, $q$ is the coupling factor between p- and g- modes, and $\epsilon_g$ is the g-mode phase offset. To identify rotationally-split mixed modes, \citet{Mosser_2015} introduced stretched periods $\tau$, defined by
\begin{equation}\label{eqn:tau}
    \mathrm{d}\tau = \frac{1}{\zeta}\frac{\mathrm{d}\nu}{\nu^2}.
\end{equation}
The mixed-mode stretching factor $\zeta(\nu)$ \citep{Deheuvels_2012,Goupil_2013} describes how much \textit{observed} period spacing deviate from the asymptotic value of $\Delta\Pi_1$. It is formally defined by the following\footnote{This description of $\zeta$ follows the representation introduced by \citet{Hekker_2017} and is equivalent to the original representation by \citet{Mosser_2015}.}:
\begin{equation}\label{eqn:zeta}
\zeta(\nu) = \bigg[1 + q\frac{\Delta\Pi_1\,\nu_{n_p}^2}{\Delta\nu_p}\frac{1}{q^2\cos^2(\pi\frac{\nu_m-\nu_{n_p}}{\Delta\nu_p}) + \sin^2(\pi\frac{\nu_m-\nu_{n_p}}{\Delta\nu_p})}   \bigg],
\end{equation}
where $\Delta\nu_p$ is the frequency difference between consecutive radial orders $n_p$ and $n_p + 1$. The coordinate transformation from $\nu$ to $\tau$ thus involves the numerical integration of Equation \ref{eqn:tau}, which is dependent on the choice of $\Delta\Pi_1$ and $q$.

\subsection{Bayesian Optimization}\label{section:bayesopt}

\begin{figure}
    \centering
	\includegraphics[width=1.\columnwidth]{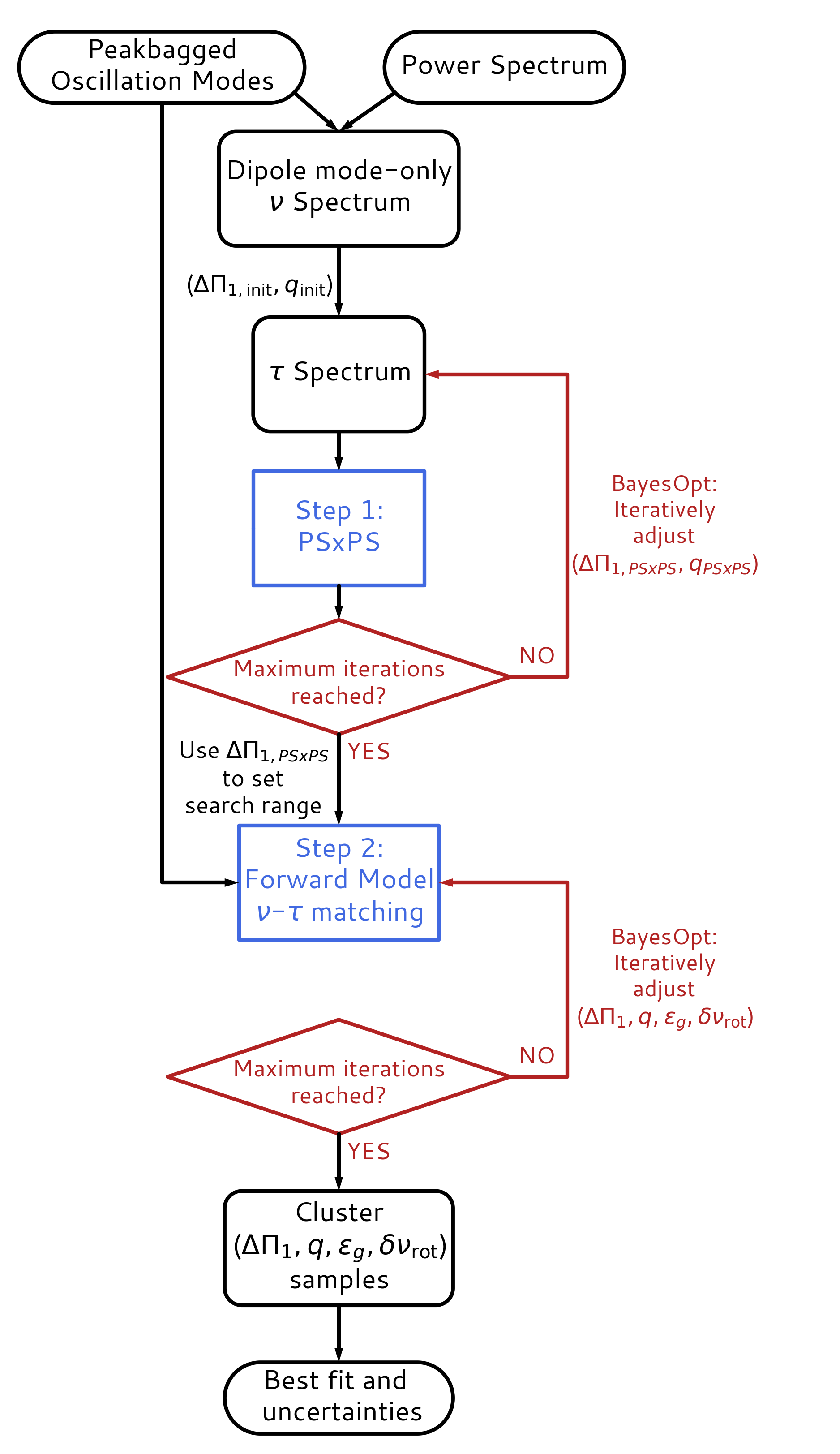}
    \caption{A flowchart detailing the estimation of mixed-mode parameters of in this work. Bayesian optimization (BayesOpt) is used in the two procedures shown in blue. The first optimization step iteratively determines estimates of the asymptotic dipole mode period spacing, $\Delta\Pi_{\mathrm{1, PSxPS}}$, from the power spectrum of the stretched period ($\tau$) spectrum (PSxPS) method. Next, $\Delta\Pi_{\mathrm{1, init}}$ initializes the search range for the second optimization step that estimates the four-parameter solution ($\Delta\Pi_1, q, \epsilon_g, \delta\nu_{\mathrm{rot}}$) from matching both $\tau$ and $\nu$ between the model and peakbagged modes.}
    \label{fig:flowchart}
\end{figure}

Bayesian Optimization \citep[hereafter BayesOpt]{Mockus_1975} comprises optimization algorithms that learn to locate the global optimum of a given objective using exploration methods guided by Bayesian statistics. The overall strategy of BayesOpt is to explore the parameter space associated with an objective within a relatively small number of iterations, making BayesOpt useful for optimizing over high-dimensional spaces when the computational budget is limited. To do this, BayesOpt acts as a black-box optimizer, meaning that it requires only the outputs from an objective function to carry out optimization and is thus agnostic to the analytic form of that function. This property of BayesOpt makes it useful for optimizing over multi-modal parameter spaces in which the use of gradient-based optimizers like gradient descent or Newton's method (e.g., \citealt{Jones_2001}) is difficult. Furthermore, it adds flexibility in defining arbitrary and potentially non-differentiable objective functions.

There are three key procedures to a standard BayesOpt algorithm:
\begin{enumerate}
\item \textbf{Creating a surrogate model.}. Given a small number of candidate points, $x_{\mathrm{cand}}$, and the objective evaluated at those points, $f(x_{\mathrm{cand}})$, a Gaussian process models a smooth approximation of the objective (${g}$). Rather than needing to evaluate $f(x)$ directly (which can be expensive), the Gaussian Process acts as a `surrogate model' whose function approximation $g(x)$ is inexpensive to evaluate and can provide interpolated estimates across the parameter space based only a subset of points.

\item \textbf{A new selection of $x_{\mathrm{cand}}$}. \MHupdate{The optimal solution from $g(x)$ in this iteration is identified as $x_{\mathrm{cand}}$. Since $g(x)$ is a smooth, continuous approximation, $x_{\mathrm{cand}}$ may adopt any value within the domain of $f$. An \textit{acquisition function} motivated by Bayesian statistics is calculated based on $f(x_{\mathrm{cand}})$ and determines the next promising regions in the parameter space to query. The acquisition function balances between exploration and exploitation to increase the likelihood of discovering new, optimal solutions in consecutive BayesOpt iterations.}

\item \textbf{Update and iterate.} With an increasing number of iterations, \MHupdate{the Gaussian process has explored more regions of the parameter space, and thus the approximation $g$ models the underlying function $f$with increasing accuracy.} This subsequently leads to better choices on which region of parameter space to explore next and increases the likelihood of discovering better solutions. Add the end of the number of iterations set by the computational budget, the best solution within the parameter space across all iterations is returned.
\end{enumerate}

To summarize, BayesOpt is an iterative procedure that smoothly approximates a parameter space using a subset of candidate points, then decides where in that space to search next using a statistically-motivated acquisition function. In this work, we use a variant of BayesOpt known as trust region Bayesian optimization \citep[\texttt{TuRBO}]{Eriksson_2019} whose acquisition function is based on Thompson sampling \citep{Thompson_1933}. Further details of this algorithm is described in Appendix \ref{appendix:turbo}.

\MHupdate{In our study, we opted for the utilization of BayesOpt as the preferred optimization algorithm over other alternatives because it offers great flexibility in accommodating objective functions that may not adhere to conventional likelihood functions (see Section \ref{section:optimal_parameters}). Additionally, BayesOpt provides a straightforward and computationally efficient means of identifying optimal solutions based on fitted data, while making minimal assumptions about the underlying parameter space between different stars. This approach enables us to efficiently draw inferences from a large sample of stars in a hands-off approach that requires minimal manual intervention.}

\begin{figure}
    \centering
	\includegraphics[width=1.\columnwidth]{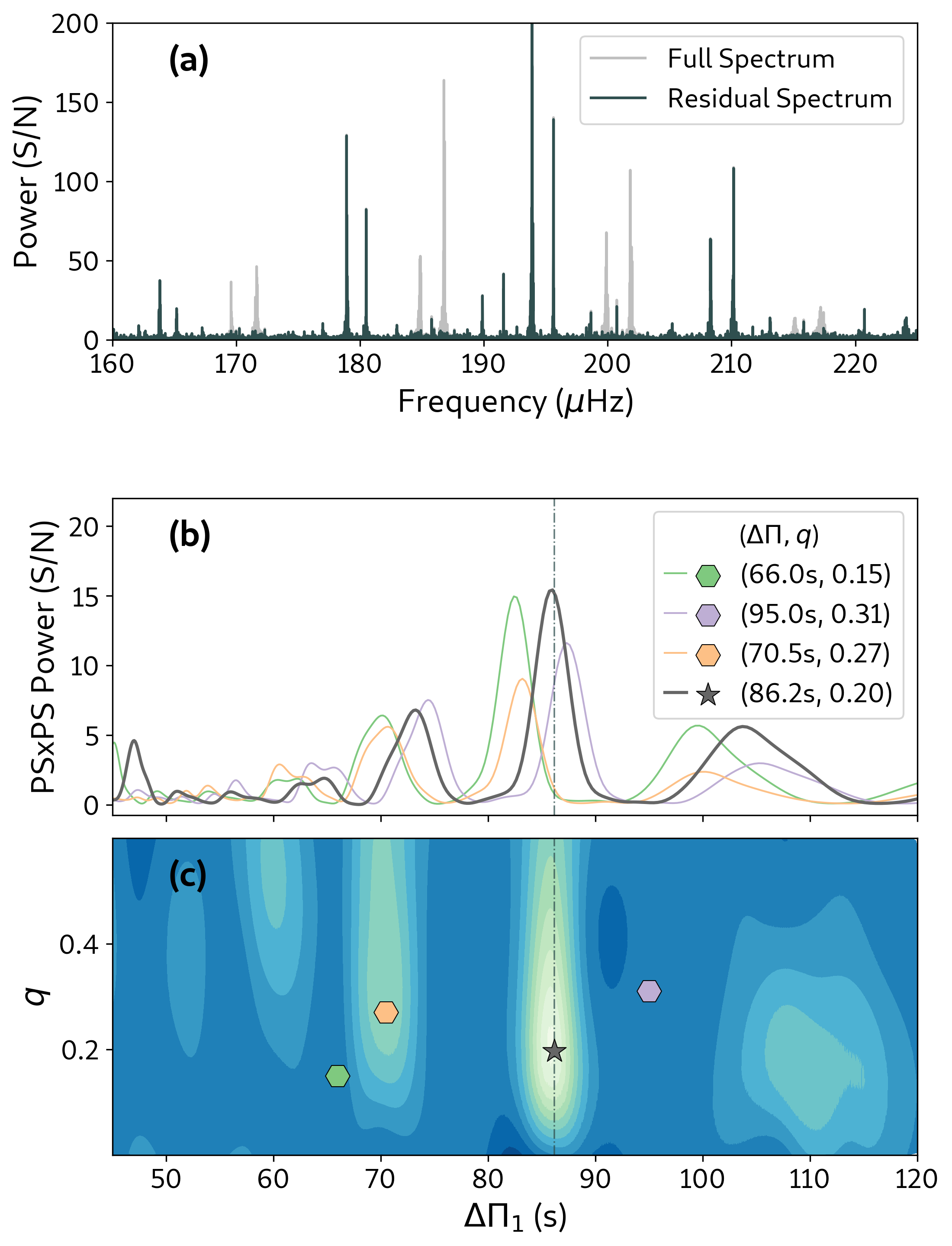}
    \caption{Optimization for the power spectrum of the stretched period spectrum for KIC 2568912. (a) The background-corrected oscillation spectrum of the star, with the $l=0,2$ modes shown in light gray. The residual spectrum is obtained by dividing the full spectrum with a template that models only the $l=0,2$ modes as described in the text. (b) The power spectrum of the $\tau$ spectrum (PSxPS) for different combinations of ($\Delta\Pi_1, q$). The ideal combination is that which maximizes the PSxPS power \textit{at that value of $\Delta\Pi_1$ itself}. \MHupdate{Here, the best solution is at $\Delta\Pi_1$ = 86.2s because the PSxPS generated using a candidate $\Delta\Pi_1$ of 86.2s peaks at that exact same value, as traced by the vertical line. In contrast, the PSxPS generated using a candidate $\Delta\Pi_1$ of 66.0s (green) has no power at $\Delta\Pi_1$ = 66.0s, which makes this solution suboptimal.} (c) The output from BayesOpt on the PSxPS, where the optimum is located at ($\Delta\Pi_1, q$) = (86.2s, 0.2). The contour represents likelihoods from the Gaussian process of the optimizer, with lighter colors indicating better ($\Delta\Pi_1, q$) solutions. }
    \label{fig:psxps}
\end{figure}

\subsection{Fitting Mixed-Mode Parameters}\label{section:optimal_parameters}
Our full procedure, as shown schematically in Figure \ref{fig:flowchart}, involves applying BayesOpt to two optimization tasks. The first is maximizing the signal in the power spectrum of the stretched period ($\tau$) spectrum \citep{Vrard_2016}, whose results are used to initialize the search range for the second optimization task. This second task is forward modelling the mixed-mode pattern by matching frequency ($\nu$) and $\tau$ values of of observed mixed-modes to those simulated using candidate values of ($\Delta\Pi_1, q, \epsilon_g, \delta\nu_{\mathrm{rot}}$). Both procedures require the background corrected oscillation spectrum of the star as well as frequencies and linewidths of individual mixed-modes \textemdash{} all of which can be provided by existing peakbagging algorithms \citep{Kallinger_2019, Corsaro_2020}.

\subsubsection{Optimizing the power spectrum of the stretched period spectrum}
\label{section:PSxPS_opt}
For each power spectrum, we perform a change of coordinate from frequency ($\nu$) to stretched period ($\tau$). To observe periodic behaviour only from the dipole mode pattern, we divide the oscillation spectrum with a template containing excess power only surrounding the central frequency of each non-dipolar mode. 
An example of the residuals in the power spectrum after dividing by the template is shown in Figure \ref{fig:psxps}a.

Next, we apply Eqns. \ref{eqn:tau}-\ref{eqn:zeta} to the residual spectrum using initial guesses ($\Delta\Pi_{\mathrm{1, init}}$, $q_{\mathrm{init}}$). We estimate $\Delta\Pi_{\mathrm{1, init}}$ using a simple linear relation:
\begin{equation} \label{eqn:scaling_dpi}
    \Delta\Pi_{\mathrm{1, init}} \mathrm{(s)} = 60 + 1.7\,\Delta\nu,
\end{equation}
which approximates the tight $\Delta\Pi_1-\Delta\nu$ relation shown by the majority of low-mass stars that are ascending the red giant branch with electron degenerate cores \citep{Vrard_2016, Deheuvels_2022}. Meanwhile, we set $q_{\mathrm{init}}$ to 0.2. The choice of ($\Delta\Pi_{\mathrm{1, init}}$, $q_{\mathrm{init}}$) does not strongly affect the outcome of the optimization due to the broad search range we apply over these two values in this step. The optimization objective is formulated as follows:
\begin{equation}\label{eqn:psxps_opt}
\begin{aligned}
\max_{\Delta\Pi_1, q} \quad & P_\mathrm{PSxPS}(\Delta\Pi_1),\\
\textrm{where} \quad & 0.8\,\Delta\Pi_{\mathrm{1, init}}\leq\Delta\Pi_1< 1.2\,\Delta\Pi_{\mathrm{1, init}},\\
  & 0 \leq q < 0.8,   \\
\end{aligned}
\end{equation}
where $P_\mathrm{PSxPS}(\Delta\Pi_1)$ is the power contained in the power spectrum of the stretched period spectrum at a candidate value of $\Delta\Pi_1$. For this task, we perform 2,000 iterations of BayesOpt and use the best solution found, ($\Delta\Pi_{\mathrm{1, PSxPS}}$, $q_{\mathrm{PSxPS}}$), for the next step of our method in this work. \MHupdate{The choice of 2,000 iterations was determined empirically to be the number of iterations that guarantees a good solution is identified within the domain of the ($\Delta\Pi_{\mathrm{1}}$, $q$). Although this implies that 2,000 instances of stretched-period spectra have to be calculated, the computational burden is light because it requires a re-determination of $\zeta$ that contains an interpolation in $\nu$ for efficiency (Appendix \ref{appendix:zeta}) and does not require explicitly solving for the mixed-mode frequencies (Eqns. \ref{eqn:tau} and \ref{eqn:zeta}). Moreover, the only information that is accumulated across BayesOpt iterations are the $\sim10^3$ points in the parameter space that have been sampled and the optimizer's Gaussian process model. The PSxPS optimization step thus typically takes only 30-60 seconds per star and is parallelizable.}

An example of the BayesOpt outcome from this step, as well as a visualization of the resulting PSxPS, is shown in Figures \ref{fig:psxps}b-c.
Note that while this optimization procedure currently focuses on red giant branch stars, it can easily be extended to red clump stars by extending the range of $\Delta\Pi_1$ that is searched over.

\begin{figure*}
    \centering
	\includegraphics[width=2.\columnwidth]{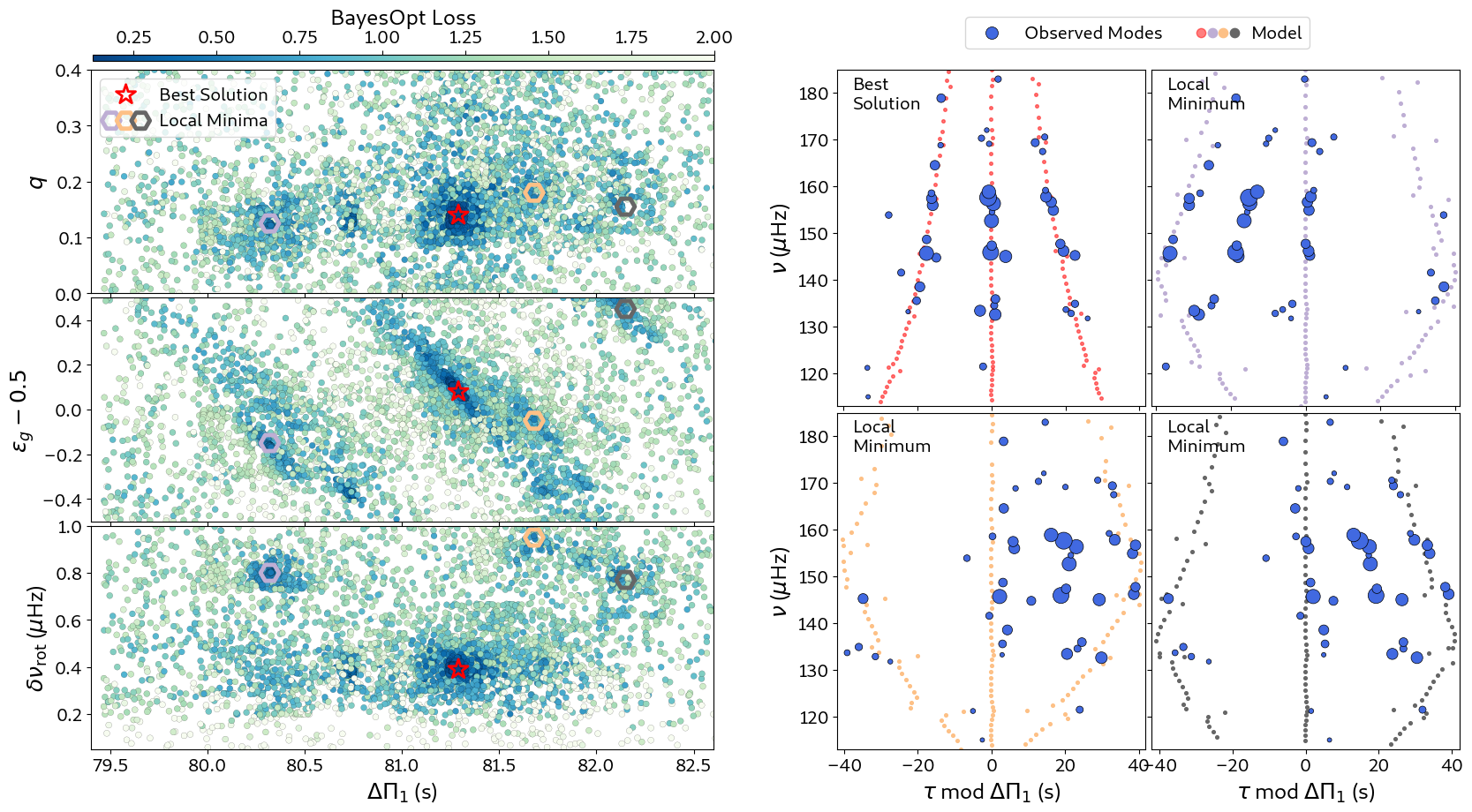}
    \caption{Bayesian optimization for the fitting of KIC 3973247. (\textit{Left}) Samples of the optimizer in the joint ($\Delta\Pi_1, q, \epsilon_g, \delta\nu_{\mathrm{rot}}$) space, shown in three separate 2D slices. Darker points correspond to better fits (lower loss). Highlighted points indicate solutions whose fits are shown in the right panel. (\textit{Right}) Example fits identified from four minima identified by the optimizer from the the joint ($\Delta\Pi_1, q, \epsilon_g, \delta\nu_{\mathrm{rot}}$) space, of which one is the global minimum or best fit. In each sub-panel, blue points correspond to observed mixed dipole mode frequencies, while the other underlying points correspond to simulated mode frequencies.} 
    \label{fig:ech_fit}
\end{figure*}

\begin{figure*}
    \centering
	\includegraphics[width=2.\columnwidth]{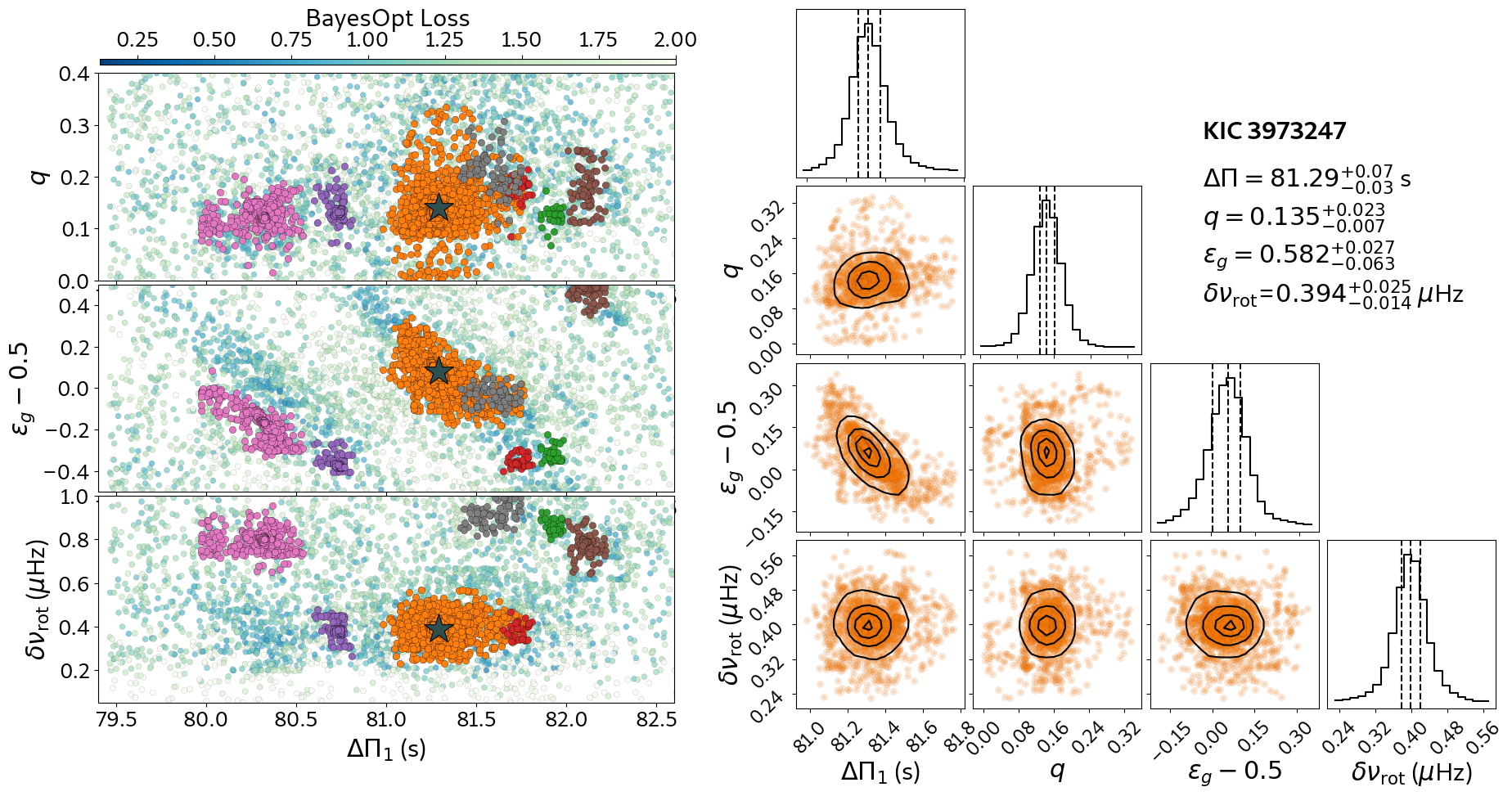}
    \caption{Estimating the uncertainties for the fitting of KIC 3973247. (\textit{Left}) Samples from the optimization, similar to that shown in Figure \ref{fig:ech_fit}. The highlighted point indicates the best fit (global optimum). The various colors indicate clusters of low-loss samples that are identified by DBSCAN. (\textit{Right}) Joint distribution of the cluster containing the best fit. The lines in each parameter's 1D histogram are the distribution median and the 67\% highest density interval (HDI). We report the HDI as 1$\sigma$ uncertainties for each fit in this work. } 
    \label{fig:clustering}
\end{figure*}

\subsubsection{Forward modelling mixed-mode mode frequencies and stretched periods}\label{section:forward_modelling}
For each observed dipole mixed-mode, we seek to find a best match of their frequencies ($\nu_{\mathrm{obs}}$) and their corresponding stretched periods ($\tau_{\mathrm{obs}}$) to those from a template, which is performed by iterating over the joint $(\Delta\Pi_1, q, \epsilon_g, \delta\nu_{\mathrm{rot}})$ parameter space. For each candidate combination of these four parameters, we generate a set of synthetic dipole mixed-modes with frequencies $\nu_s = [\nu_{\mathrm{s,m=0}}, \nu_{\mathrm{s,m=-1}}, \nu_{\mathrm{s,m=+1}}]$ and stretched periods $\tau_s=[\tau_{\mathrm{s,m=0}}, \tau_{\mathrm{s,m=-1}}, \tau_{\mathrm{s,m=+1}}]$. Here, $\tau_{s}$ is the interpolation of the $\tau$ spectrum at $\nu_{s}$ as described in Appendix \ref{appendix:zeta}, and $m$ corresponds to the azimuthal number corresponding to rotational splitting. Note that candidate values of $\Delta\Pi_1$ and $q$ that are used to construct $\tau_{\mathrm{obs}}$ are also used to construct $\tau_{\mathrm{s}}$.

\MHupdate{For each candidate value of $(\Delta\Pi_1, q, \epsilon_g, \delta\nu_{\mathrm{rot}})$, we build a k-dimensional tree\footnote{Here, $k=2$ is the dimensions of the space for which the distances are computed.} for the generated values of $(\nu_{\mathrm{s}}, \tau_{\mathrm{s}})$ using the implementation from \texttt{scikit-learn} version 1.1.2 \citep{scikit-learn}\footnote{\texttt{sklearn.neighbors.KDTree}}. For each $(\nu_{\mathrm{obs}}, \tau_{\mathrm{obs}})$ pair, we query the k-dimensional tree 
to find the $(\nu_{\mathrm{s}}, \tau_{\mathrm{s}})$ pair that is queried point's nearest neighbor in the 2D space. The Euclidean distance of the nearest neighbour match between $(\nu_{\mathrm{obs}}, \tau_{\mathrm{obs}})$ and $(\nu_{\mathrm{s}}, \tau_{\mathrm{s}})$ is the cost that BayesOpt seeks to minimize. Formally, we describe the BayesOpt objective as follows:}
\begin{equation}\label{eqn:freq-tau_matching}
\begin{aligned}
\min_{\Delta\Pi_1, q, \epsilon_g, \delta\nu_{\mathrm{rot}}} \quad & \mathrm{median}(\mathcal{D}_{\mathrm{KD}}[(\widehat{\nu}_{\mathrm{obs}}, \widehat{\tau}_{\mathrm{obs}}), (\widehat{\nu}_{\mathrm{s}}, \widehat{\tau}_{\mathrm{s}})]),\\
\textrm{where} \quad & \widehat{\nu} = (\nu-\nu_{\mathrm{max}})/\Delta\nu,\\
  & \widehat{\tau} = \tau/\Delta\Pi_1,   \\
  & | \Delta\Pi_1 - \Delta\Pi_{\mathrm{1, PSxPS}}|/\Delta\Pi_{\mathrm{1, PSxPS}} \leq0.02\\
  & 0.05\leq q < 0.4\\
  & -0.5\leq \epsilon_g - 0.5 < 0.5\\
  & 0.05\,\mu\mathrm{Hz}\leq\delta\nu_{\mathrm{rot}} < 1\,\mu\mathrm{Hz}\\
\end{aligned}
\end{equation}

Here, $\Delta\Pi_{\mathrm{1, PSxPS}}$ is the best fit to $\Delta\Pi_1$ previously determined from the PSxPS (Section \ref{section:PSxPS_opt}), $\mathcal{D}_{\mathrm{KD}}$ is the Euclidean distance as measured by the k-dimensional tree  and $\Delta\nu$ is the large frequency separation. Note that we use normalized values for both $\tau$ and $\nu$ in Equation \ref{eqn:freq-tau_matching} to prevent either from influencing the distance computation too heavily due to differences in scale. For each star in this study, we perform 7,500 iterations of BayesOpt and report the best value of ($\Delta\Pi_1$, $q$, $\epsilon_g, \delta\nu_{\mathrm{rot}}$) identified from the optimization. 
\MHupdate{This number of iterations was determined empirically to provide a sufficiently good exploration of the 4D parameter space such that good fits could be consistently identified across our dataset. At the same time, the need to minimize computational cost was considered when determining the appropriate number of iterations. The optimization process for each star takes approximately 10-15 minutes, resulting in a total processing time of around 1-2 days for the entire dataset when parallelization is employed.}

This formulation assumes that all three rotational components are present in the mixed-modes, which simplifies the optimization task. If three rotational components are not observed in the real data, the BayesOpt algorithm typically identifies a fit close to globally optimal solution, but with a $\delta\nu_{\mathrm{rot}}$ that is an integer multiple of the correct value. Such cases are easily identifiable by visual inspection of the stretched \'echelle diagram. For these, we re-run the optimization with the correct number of rotational components. \MHupdate{Further discussion of this assumption and its corresponding limitation is discussed in Section \ref{section:limitations} and Appendix \ref{appendix:triplet}.}

\begin{figure}
    \centering
	\includegraphics[width=1.\columnwidth]{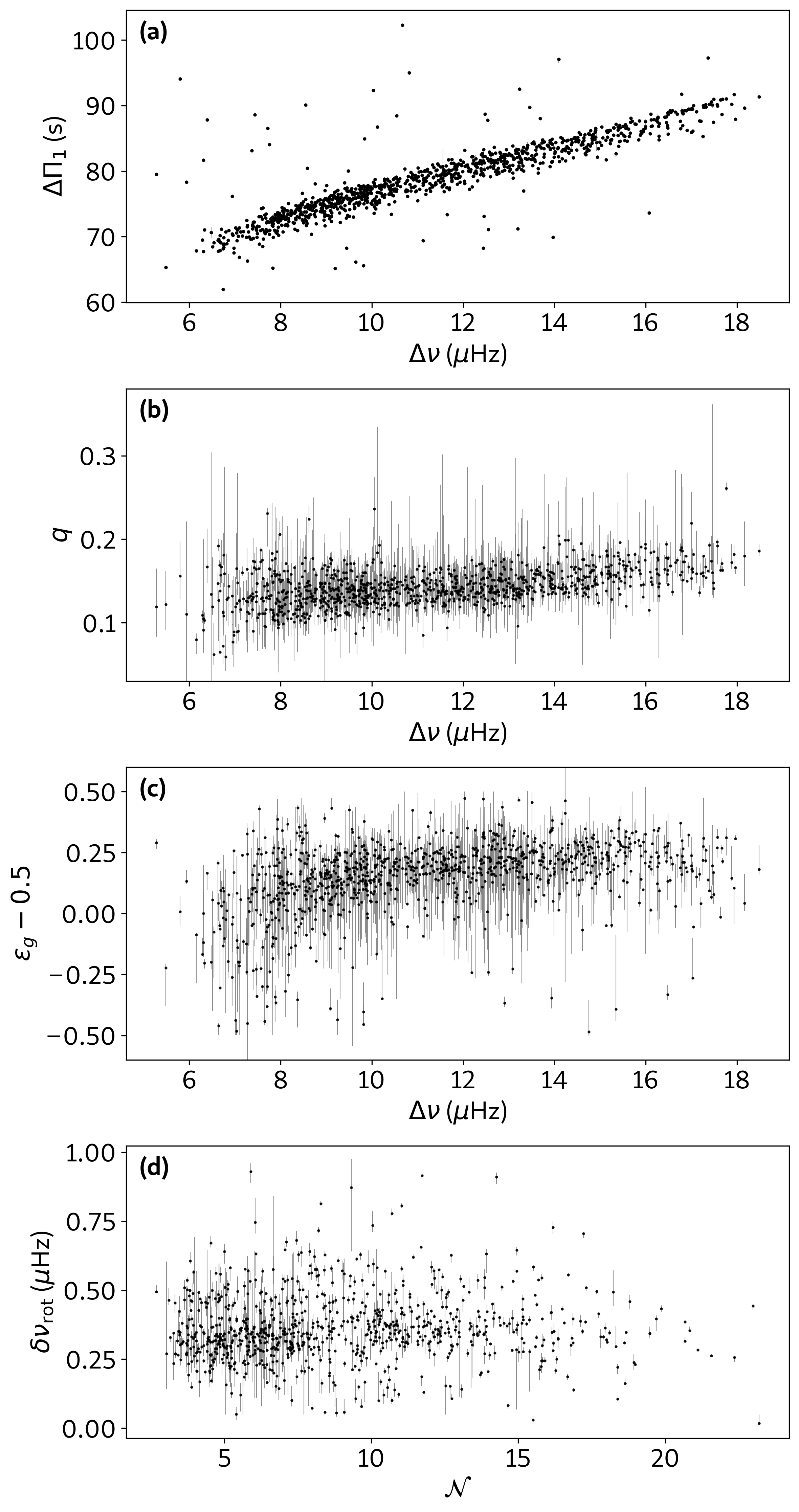}
    \caption{Mixed-mode parameters for the 1,068 \textit{Kepler} red giants from the \citet{Gehan_2018} study.  (\textit{a}) The $\Delta\Pi_1$-$\Delta\nu$ relation as previously investigated by \citet{Mosser_2014} and \citet{Vrard_2016}. (\textit{b}) The $\Delta\nu-q$ relation, previously demonstrated by \citet{Mosser_2017}. (\textit{c}) The $\epsilon_g$-$\Delta\nu$ relation \citep{Mosser_2018}. (\textit{d}) The observed mixed-mode rotationally frequency splitting, $\delta\nu_{\mathrm{rot}}$, as a function of the mixed-mode density $\mathcal{N} = \Delta\nu/(\Delta\Pi_1\cdot\nu_{\mathrm{max}}^2)$ \citep{Mosser_2012, Gehan_2018}.}
    \label{fig:summary}
\end{figure}

\begin{figure}
    \centering
	\includegraphics[width=1.\columnwidth]{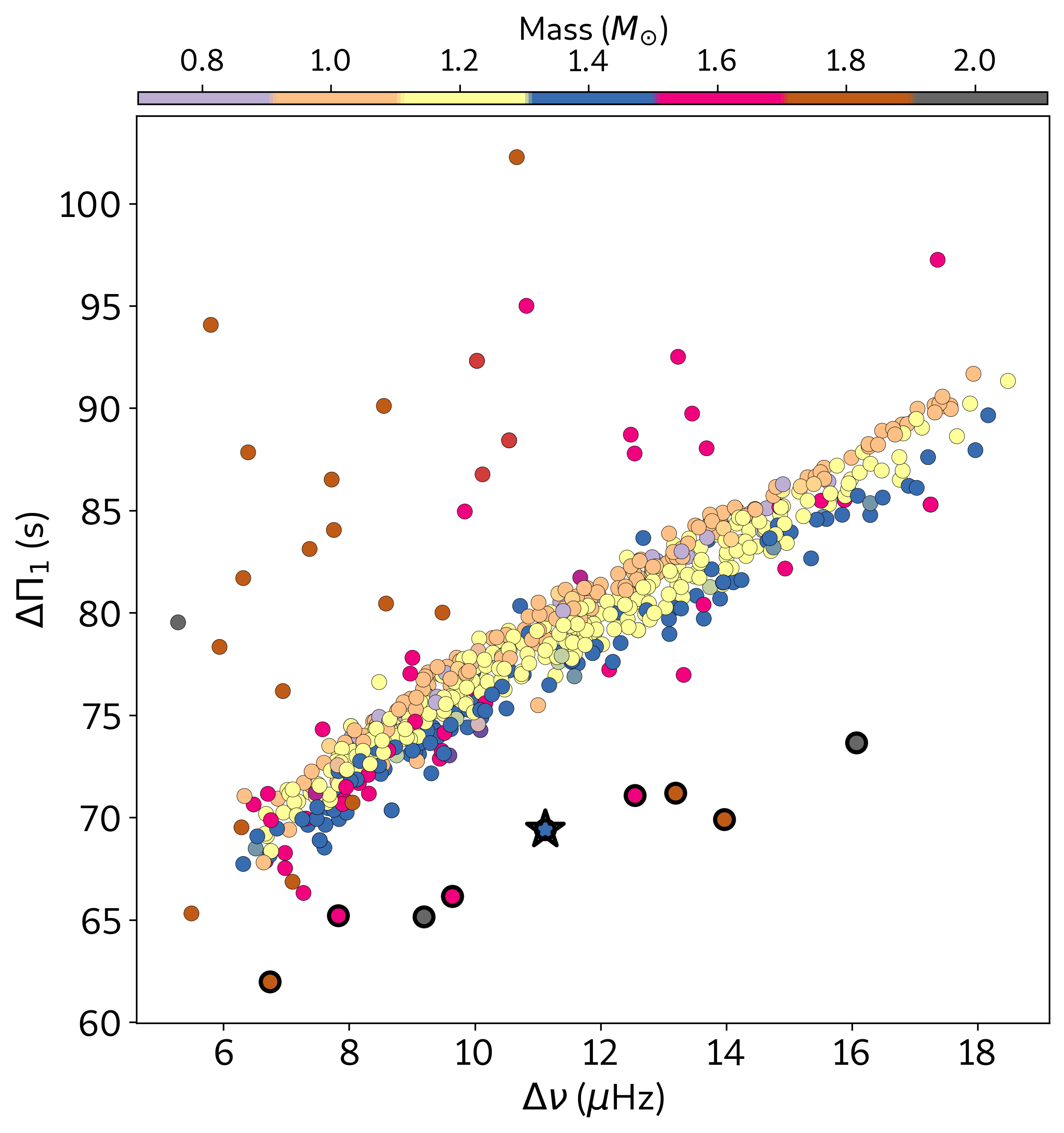}
    \caption{\MHupdate{The $\Delta\Pi_1-\Delta\nu$ relation for 684 stars with atmospheric parameters from SDSS/APOGEE DR17 across a range of stellar mass bins.} Highlighted stars are those that with a large vertical distance ($>\,$6$\,$s) from the `degenerate sequence' \citep{Deheuvels_2022}. The target marked with a star symbol is KIC 4350501, a known young alpha-rich star \citep{Martig_2015}. }
    \label{fig:dpi_cn}
\end{figure}


\begin{figure*}
    \centering
	\includegraphics[width=2.\columnwidth]{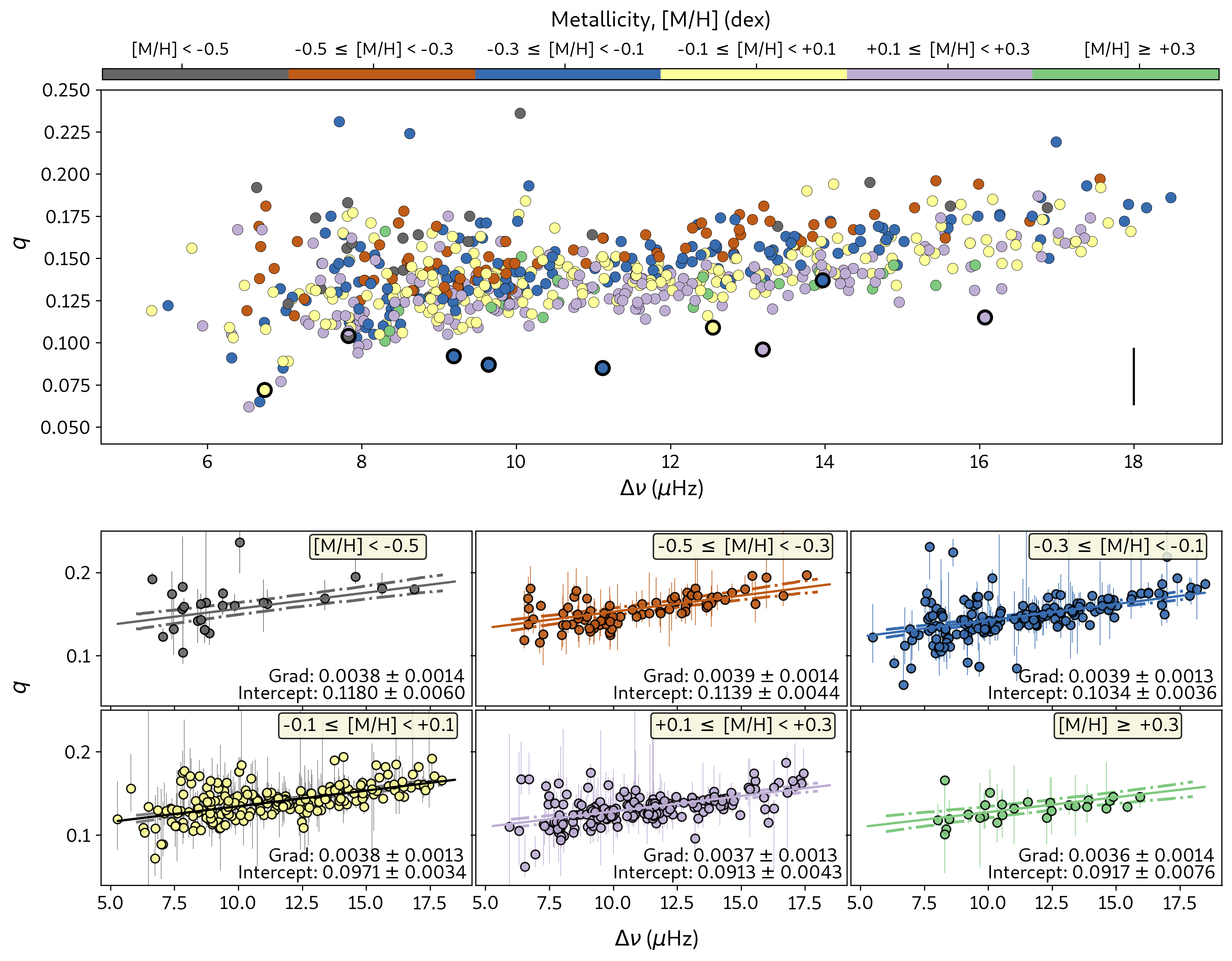}
    \caption{The $\Delta\nu-q$ relation for red giants across different metallicity ([M/H]) bins. The top panel shows all 684 targets with abundances from APOGEE DR17. Highlighted points are stars below the $\Delta\Pi_1$-$\Delta\nu$ degenerate sequence in Figure \ref{fig:dpi_cn}a. The bottom panels show the $\Delta\nu-q$ relation for stars within specific metallicity bins. A linear relation is fit to the targets in each bin using a hierarchical linear model (see text), where the intervals indicate 95\% highest density regions from the posterior predictive distribution of the fit.} 
    \label{fig:dnu_q}
\end{figure*}

\begin{figure*}
    \centering
	\includegraphics[width=2.\columnwidth]{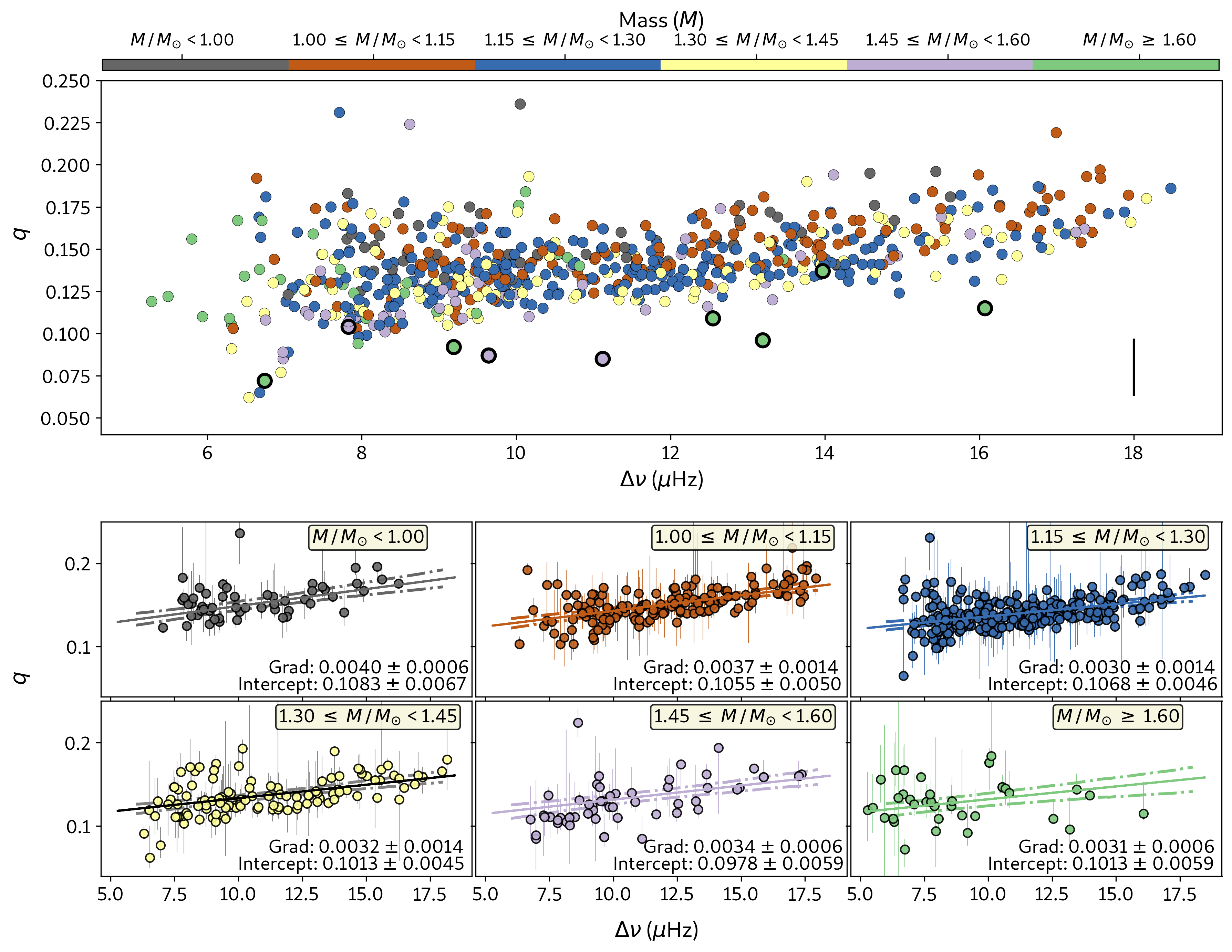}
    \caption{Same as Figure \ref{fig:dnu_q}, but for bins of stellar mass.} 
    \label{fig:dnu_q_mass}
\end{figure*}

\begin{figure}
    \centering
	\includegraphics[width=1.\columnwidth]{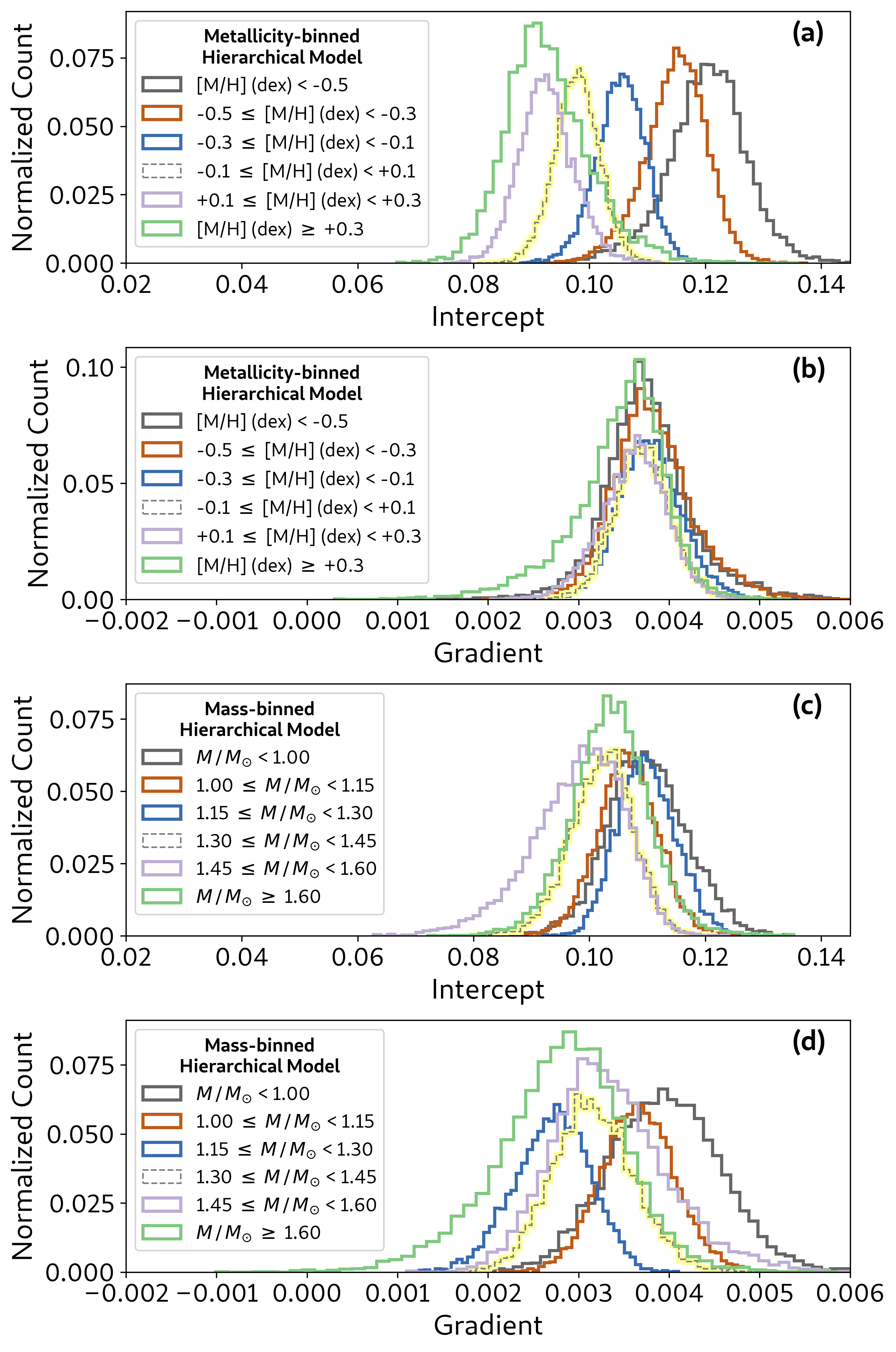}
    \caption{Intercepts and gradients to the $\Delta\nu-q$ relation from the posterior predictive samples, binned in metallicity (panels \textit{a-b}) and stellar mass (panels \textit{c-d}). The relation was fitted separately for mass and metallicity using hierarchical linear models, with the posterior distribution estimated using Markov Chain Monte Carlo techniques implemented in \texttt{pymc3} \citep{pymc3}.} 
    \label{fig:intercepts}
\end{figure}

\begin{figure}
    \centering
	\includegraphics[width=1.\columnwidth]{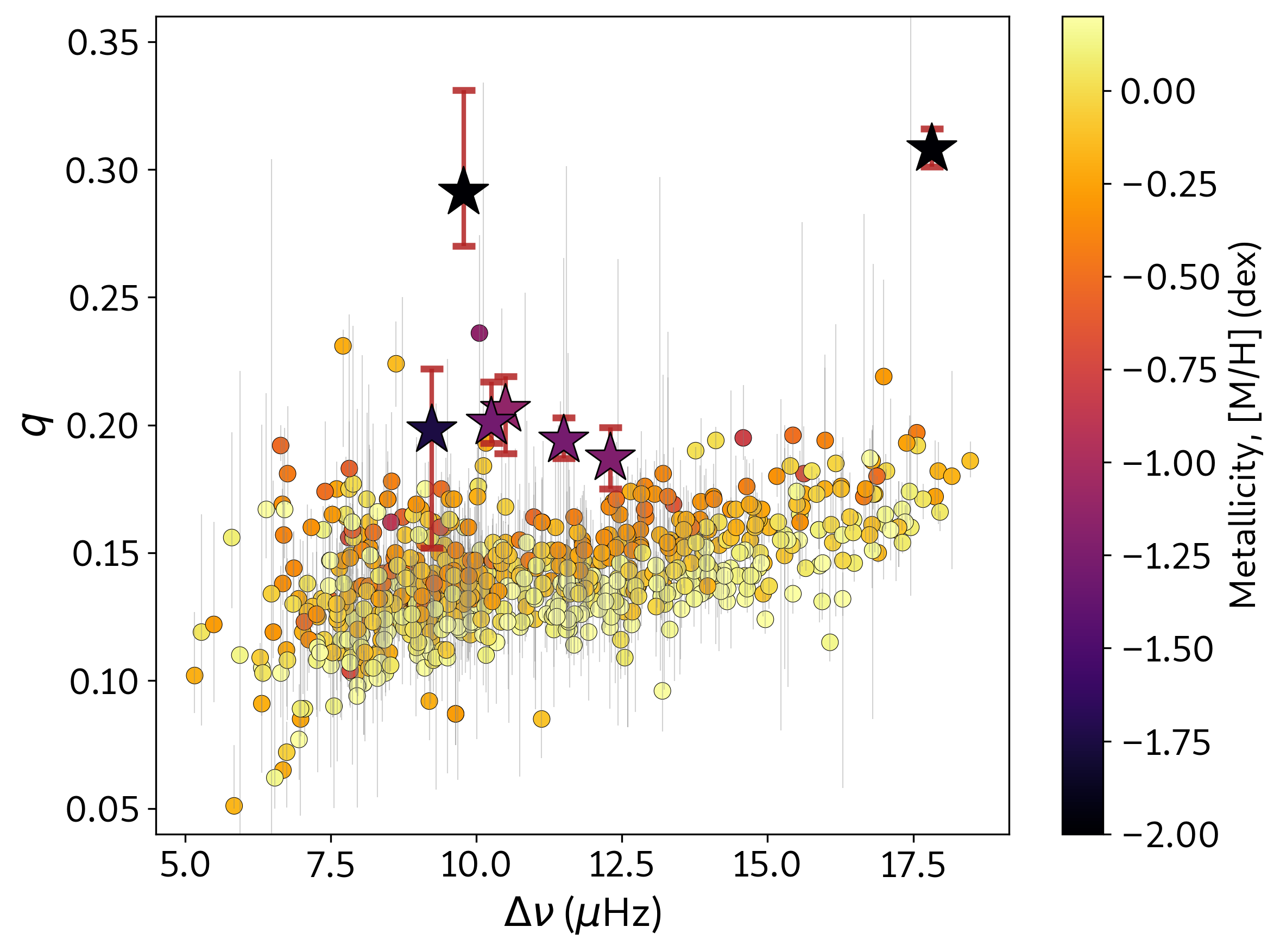}
    \caption{Metallicity dependence of the $\Delta\nu-q$ sequence in our sample. Highlighted are 7 new, metal-poor ([M/H] $<$ -1.0 dex) \textit{Kepler} targets not present in the \citet{Gehan_2018} sample, with parameters tabulated in Table \ref{table:metal-poor}.}
    \label{fig:metal_poor}
\end{figure}

\subsection{Reported Uncertainties}\label{section:uncertainties}.
The tendency of the optimization algorithm to densely sample regions near minima in the joint parameter space (see Figure \ref{fig:ech_fit}) provides a convenient way of quantifying the uncertainty of our fits. In particular, we determine how compact sampled points are clustered around the best-fit solution. To cluster the data, we apply the DBSCAN algorithm\footnote{\texttt{sklearn.cluster.DBSCAN}} \citep{Ester_1996} as implemented in \texttt{scikit-learn} version 1.1.2 \citep{scikit-learn}. We use the default settings with the exception of the parameter $\epsilon_{\mathrm{clust}}$ that determines the maximum extent of the neighbourhood surrounding a point to consider as part of a cluster. Following \citet{Rahmah_2016}, we estimate the optimal $\epsilon_{\mathrm{clust}}$ by calculating the Euclidean distance of each point to its nearest neighbour. By sorting the points in order of increasing nearest neighbour distance, we create a clustering distance profile that is commonly known also as an `elbow curve'. We identify the distance at which the curve as the optimal value for $\epsilon_{\mathrm{clust}}$. We systematically determine where this happens for each curve using \texttt{kneebow} \citep{kneebow}.

The outcome of the DBSCAN algorithm produces a series of clusters corresponding to sampled points around local and global minima as shown in Figure \ref{fig:clustering}. Because we are mainly interested on the best-fitting solution, we report the 67\% highest density interval of the cluster containing the global minimum as 1$\sigma$ uncertainties.


\subsection{Data}\label{section:data}

For our analysis in this study, we extract the mixed-mode parameters of 1,183 stars analyzed in the work by \citet{Gehan_2018}, which is a sample sufficiently representative of low-luminosity red giants within the \textit{Kepler} mission. Using light curves from the KEPSEISMIC database \citep{KEPSEISMIC}, we measured global seismic parameters $\Delta\nu$ and $\nu_{\mathrm{max}}$ with the asteroseismic data analysis pipeline described by \citet{Themessl_2020} and subsequently extracted mode frequencies by `peakbagging' using the peak detection algorithm described by \citet{Garcia_2018}. The peakbagging algorithm nominally identifies $l=0,2$ modes, but it also detects other significant peaks in the oscillation spectrum. 
\MHupdate{The $l=0,2$ modes were identified by fitting ridges in the echelle diagram calculated using the asymptotic relation to the observed frequencies \citep{Mosser_2011, Themessl_2020}. The parameters are refined from initial scaling relation values \citep{Mosser_2011} iteratively to find a best-fitting set of parameters from which the most likely candidate frequencies were chosen as the $l=0,2$ modes by taking into account their amplitude and linewidths.} 

\MHupdate{Therefore, the input modes to the forward modelling BayesOpt step are all other significant peaks that are not $l=0,2$ modes, which are mostly dipole mixed-modes. Detected $l=3$ modes in the spectrum may, however, present as outliers when performing the $\nu-\tau$ matching in Section \ref{section:forward_modelling}. Such an occurrence does not strongly hinder our approach because the BayesOpt objective in Equation \ref{eqn:freq-tau_matching} uses median distances, which are robust towards the presence of outliers.
}


We successfully peakbagged 1,140 red giants from the initial sample of 1,183. From these 1,140 stars, we found 1,068 whose best ($\Delta\Pi_1, q, \epsilon_g, \delta\nu_{\mathrm{rot}}$) BayesOpt solution yield a forward model that fits the observed mixed-mode pattern with $\mathcal{D}_{\mathrm{KD}}\leq0.05$ and produces meaningful clusters in the 4D parameter space. 
The data products from our analysis, which include the global seismic parameters, background corrected power spectra, the list of detected frequencies and $l=0,2$ mode identifications for each red giant are made openly available at \dataset[10.5281/zenodo.7888633]{\doi{10.5281/zenodo.7888633}}. Our analysis code is hosted in the Bayesian Optimisation for the Characterisation of Mixed-Modes repository at \url{https://github.com/jsk389/BOChaMM/}.

\begin{deluxetable*}{ccccccc}[t]
\tablenum{1}
\tablecaption{Mixed-mode parameters of 1,068 \textit{Kepler} red giants from the \citet{Gehan_2018} sample. A reported value of $\delta\nu_{\mathrm{rot}} = 0$ implies only the $m=0$ mixed-mode component was detected. Whenever available, we report bulk metallicities of the star from APOGEE DR17 \citep{Abdurrouf_2022}, and subsequently report asteroseismic scaling masses using \texttt{asfgrid} \citep{Sharma_2016, asfgrid_2022}. 
The full version of this table is available in a machine-readable format in the online journal, with a portion shown here for guidance regarding its form and content. \label{table:mixed_mode_parameters}}
\tablewidth{0pt} 
\tablehead{ 
\colhead{KIC} & \colhead{$\Delta\Pi_1$} & \colhead{$q$} & \colhead{$\epsilon_g-1/2$} & \colhead{$\delta\nu_{\mathrm{rot}}$} &  \colhead{[M/H]}  & \colhead{$M$} 
\\
\colhead{} & \colhead{(s)} & \colhead{} & \colhead{} & \colhead{($\mu$Hz)} &  \colhead{(dex)} &   \colhead{($M_{\odot}$)}
}
\startdata
1027337 & $70.24^{+0.04}_{-0.07}$ & $0.077^{+0.022}_{-0.016}$ & $-0.015^{+0.100}_{-0.144}$ & $0$& $+0.21$& $1.30\pm0.08$\\
1430118 & $83.03^{+0.44}_{-0.03}$ & $0.155^{+0.022}_{-0.014}$ & $0.330^{+0.034}_{-0.288}$ & $0.279^{+0.188}_{-0.045}$& $-0.16$& $1.05\pm0.06$\\
1433803 & $80.22^{+0.27}_{-0.01}$ & $0.138^{+0.016}_{-0.019}$ & $0.039^{+0.022}_{-0.109}$ & $0.330^{+0.442}_{-0.174}$ & $+0.22$& $1.22\pm0.07$\\
1569842 & $80.71^{+0.08}_{-0.14}$ & $0.155^{+0.017}_{-0.018}$ & $0.195^{+0.105}_{-0.134}$ & $0.366^{+0.030}_{-0.027}$& $-0.28$& $0.98\pm0.06$\\
$\cdots$&$\cdots$&\hspace{10pt}$\cdots$&\hspace{8pt}$\cdots$&$\cdots$&\hspace{7pt}$\cdots$&$\cdots$\\
12834442 & $83.63^{+0.05}_{-0.04}$ & $0.148^{+0.017}_{-0.020}$ & $0.243^{+0.002}_{-0.013}$ & $0.569^{+0.041}_{-0.010}$ & -- & -- \\
12885196 & $72.56^{+0.01}_{-0.01}$ & $0.149^{+0.020}_{-0.014}$ & $0.084^{+0.010}_{-0.003}$ & $0.514^{+0.003}_{-0.007}$& $+0.16$& $1.35\pm0.08$\\
\enddata
\end{deluxetable*}

\section{Results}\label{section:results}
\subsection{Summary of mixed-mode parameters}
Figure \ref{fig:summary} shows the fitted mixed-mode parameters of the 1,068 \textit{Kepler} red giants included in the \citet{Gehan_2018} study. To investigate how these parameters vary with the other global stellar properties, we use bulk metallicity ([M/H]) and effective temperature ($T_{\mathrm{eff}}$) measurements for 684 targets with spectroscopic data from APOGEE DR17. Using these atmospheric parameters, we use model-calibrated asteroseismic scaling relations (\texttt{asfgrid}, \citealt{Sharma_2016, asfgrid_2022}) to estimate the stellar mass of each red giant. Table \ref{table:mixed_mode_parameters} tabulates the properties for each target in this study.

\subsection{Distribution of Asymptotic Dipole Mode Period Spacing}\label{section:dpi}
Figure \ref{fig:dpi_cn} shows the distribution of $\Delta\Pi_1$ as a function of stellar mass for the targets with atmospheric parameters from APOGEE in our study. As shown previously by \citet{Vrard_2016} and \citet{Deheuvels_2022}, low- to intermediate-mass ($M\apprle1.4M_{\odot}$) low-luminosity giants for stars with $\Delta\nu \apprge18\mu$Hz have converged to a common pathway in $\Delta\Pi_1$-$\Delta\nu$ space (i.e., the `degenerate sequence'). The sequence in Figure \ref{fig:dpi_cn}a shows a distinct stratification with stellar mass, corresponding to the weak mass dependence of the sequence that was predicted using models by \cite{Deheuvels_2022}. Also in agreement with models is the prevalence of more massive ($M\apprge1.6M_{\odot}$) red giants above the degenerate sequence, corresponding to the late development of electron-degenerate helium cores in such stars.

\citet{Rui_2021} and \citet{Deheuvels_2022} described the seismic identification of red giant candidates with a history of mass accretion from a stellar companion after they have developed degenerate cores. These candidates are massive red giants located below the degenerate sequence, which are also found in our results (highlighted points in Figure \ref{fig:dpi_cn}a). Of particular interest is KIC 4350501 (marked with a star symbol), which is a known young $\alpha$-rich star \citep{Martig_2015}. This supports the hypothesis describing some of these stars \textemdash{} which are peculiar from a Galactic evolution perspective \textemdash{} as products of binary mass transfer \citep{Jofre_2016}.


\subsection{Distribution of Mode Coupling Factor}\label{section:q}
The mode coupling factor $q$ determines the strength of the coupling between the p- and g- mode cavities, and therefore provides diagnostic information regarding the structure of the evanescent region between the two cavities for low-luminosity red giants \citep{Takata_2016, Mosser_2017, Hekker_2018, Pincon_2020}. In this study, we assume $q$ to be constant within the observable frequency range. This is a simplifying aspect in our bulk analysis over this large sample of red giants, and we discuss the limitations of this assumption in Section \ref{section:limitations}.

Observationally, $q$ has been shown to gradually decrease as stars with $\Delta\nu \apprle 18\mu$Hz evolve up the red giant branch, converging into an approximately linear $\Delta\nu-q$ sequence \citep{Mosser_2017}. This phenomenon has been attributed to the progressive migration of the evanescent region from the radiative to the convective zones within the star \citep{Pincon_2020}. Our measurements of this coupling factor, presented in Figures \ref{fig:summary}b and \ref{fig:dnu_q}, shows an identical sequence. 
We note that the stars previously highlighted as seismic candidates of mass transfer in Section \ref{section:dpi} appear below this sequence, and their existence in this region may be described using the same reasoning applied by \citet{Deheuvels_2022}. In particular, these stars should have the core properties of a more evolved giant (low $q$), 
but the transfer mass into their envelope from a stellar (secondary) companion results in an increase of the primary's mean density. The effect of this is keep $q$ constant while increasing $\Delta\nu$, placing such stars under the nominal $\Delta\nu-q$ sequence.

The study by \citet{Jiang_2020} \MHupdate{showed that stellar models predict a dependency of $q$ on both mass and metallicity, in the sense that varying either parameter causes vertical offsets to a star's location within the $\Delta\nu-q$ sequence. In Figures \ref{fig:dnu_q} and \ref{fig:dnu_q_mass}, we indeed observe that binning the sequence in [M/H] and stellar mass results in its stratification to a certain level. To quantify this effect for metallicity, we first bins the stars either according to their [M/H] measurements and perform a Bayesian fit of hierarchical linear models across the bins. Such a fit constrains the gradient and intercept across bins to be drawn from a common underlying distribution for which uninformative priors have been used\footnote{The prior distribution for the gradient and intercept across bins is parameterized by a mean of $\mathcal{N}(0, 100)$ and a dispersion of $|\mathcal{N}|(0, 5)$.}. We repeat the same analysis for mass bins and summarize our results in Figure \ref{fig:intercepts}. Panels (a) and (b) indicate that stars of lower metallicity will consistently sit at higher values of $q$ compared to metal-rich stars of the same $\Delta\nu$. We readily see this pattern in Figure \ref{fig:metal_poor}, which shows the influence of metallicity on the $\Delta\nu-q$ sequence without any binning. To further demonstrate the influence of metallicity, we also extract the mixed-mode parameters of an additional six metal-poor \textit{Kepler} targets that were not included in the \citet{Gehan_2018} sample. The parameters for these additional stars are tabulated in Table \ref{table:metal-poor}, with their fits to the stretched period \'echelle shown in Appendix \ref{appendix:metal-poor_ech}. From these, KIC 4671239 and KIC 8144907 are remarkably metal-poor ([M/H] $<$ -2.0 dex) with the latter's elemental abundances measured from optical high-resolution spectroscopy (Huber et al., in prep). These very metal-poor stars have
unusually large values of $q$ that span up to $0.3$, which suggests the ability of mixed-mode asteroseismology as a means of detecting metal-deficient ancient stars around our Galaxy.} 

\MHupdate{
Figures \ref{fig:metal_poor}c-d shows that there is also a dependence of $q$ on mass. This relationship is less straightforward than a vertical offset in the $\Delta\nu-q$ sequence given that the gradients of the linear fit show variations across mass bins. However, we suggest caution in interpreting these results we have treated our analysis of mass and metallicity separately, when in reality a significant degree of correlation exists between the two parameters (Appendix \ref{appendix:mass_met_corr}). Properly disengtangling the individual contributions of mass and metallicity towards $q$ will require a detailed analysis taking into account the relationship between mass and metallicity, which we leave for future work. 
}

\subsection{Distribution of g-Mode Phase Offset}

\begin{figure}
    \centering
	\includegraphics[width=1.\columnwidth]{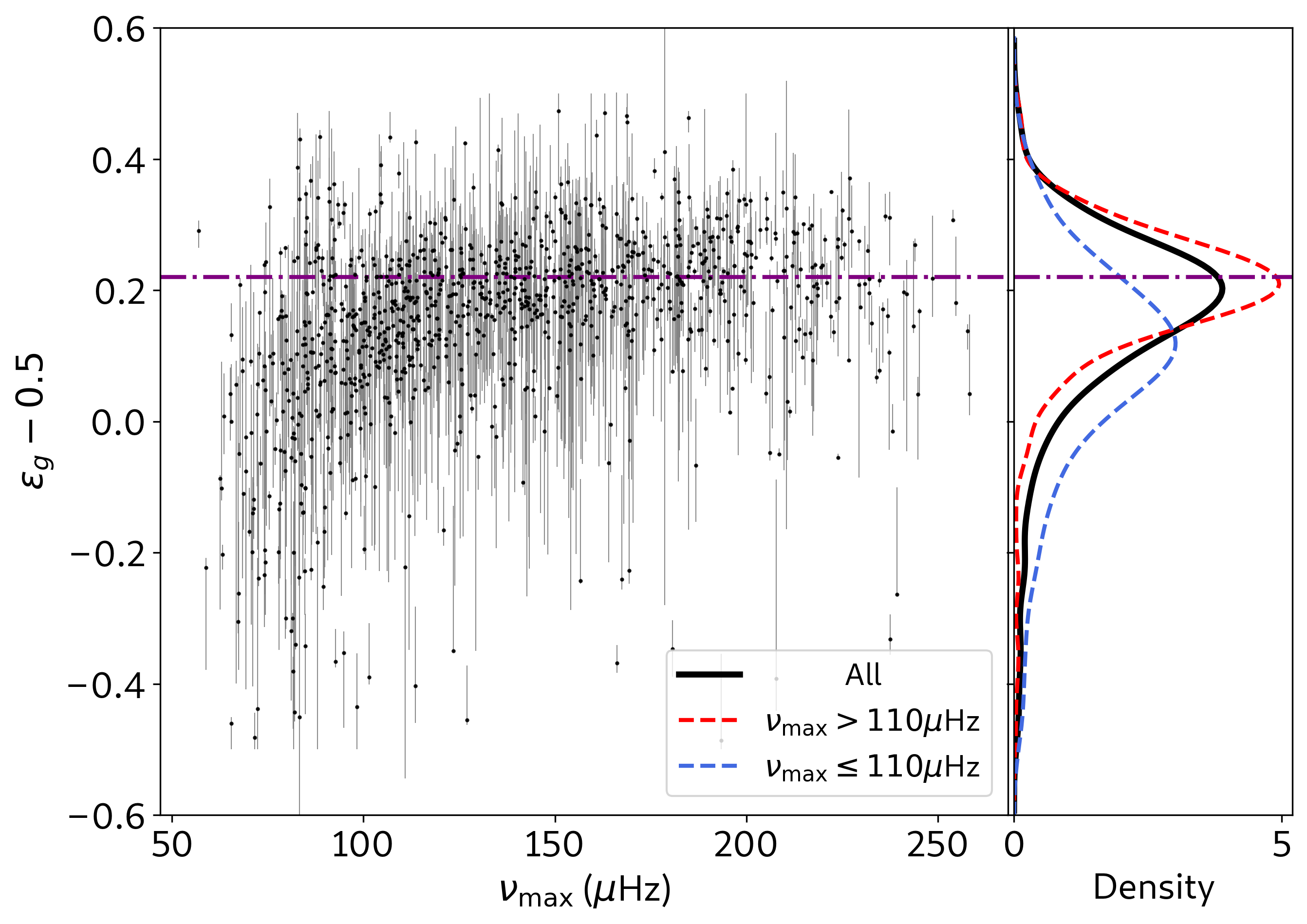}
    \caption{The distribution of $\epsilon_g$ for the \textit{Kepler} red giants in our study. The horizontal line indicates the asymptotic value of 0.22 predicted by \citet{Pincon_2019} in the weak coupling limit for stars early in their ascent of the red giant branch (RGB). Histograms are plotted separately for stars before and after 110$\mu$Hz in their evolution to highlight early RGB stars.}
    \label{fig:eg_hist}
\end{figure}

\begin{figure}
    \centering
    \includegraphics[width=1.\columnwidth]{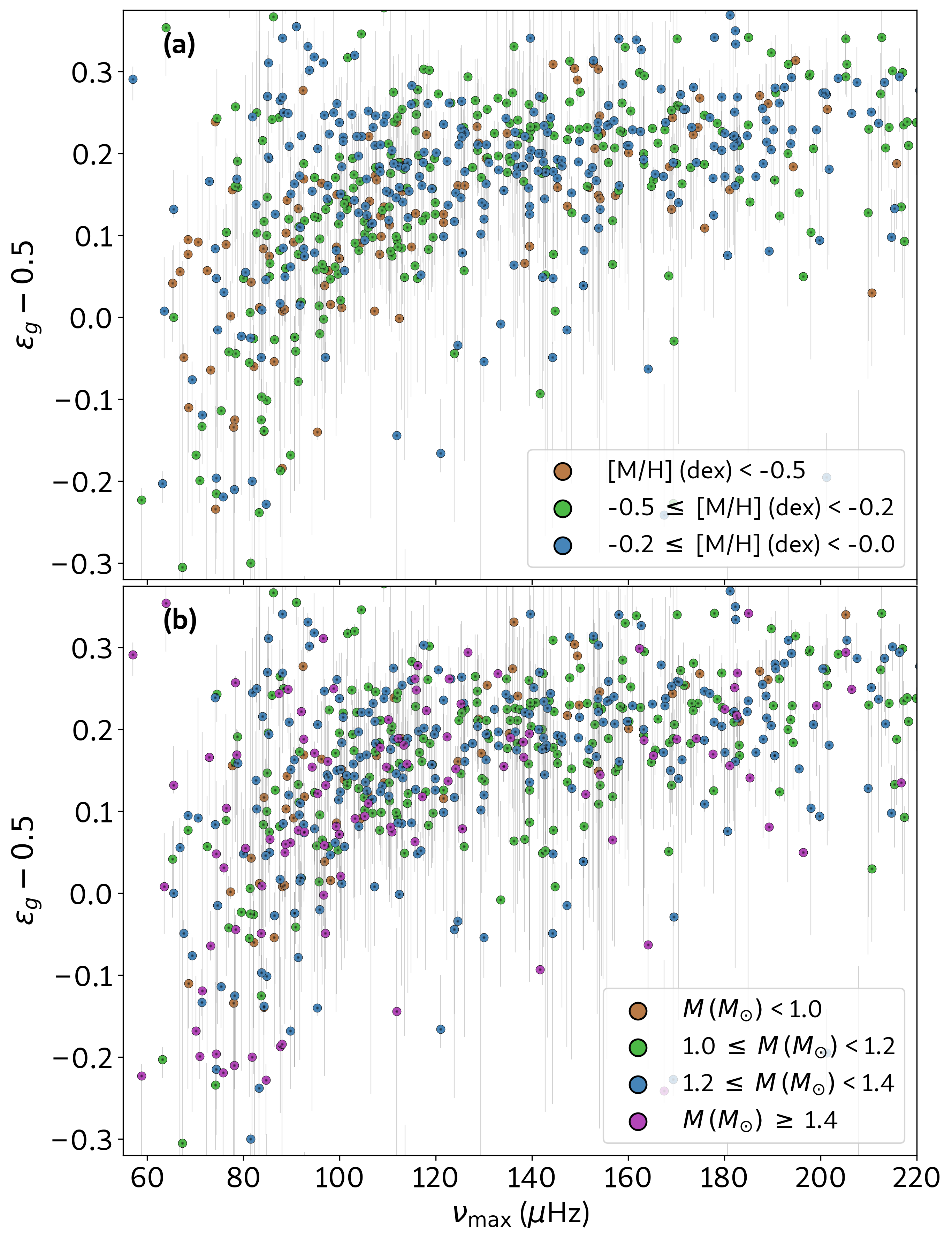}
    \caption{The evolution of $\epsilon_g$ for the stars in our sample as a function of (\textit{a}) metallicity and (\textit{b}) mass. }
    \label{fig:eg_mass_met}
\end{figure}

\begin{figure}
    \centering
    \includegraphics[width=1.\columnwidth]{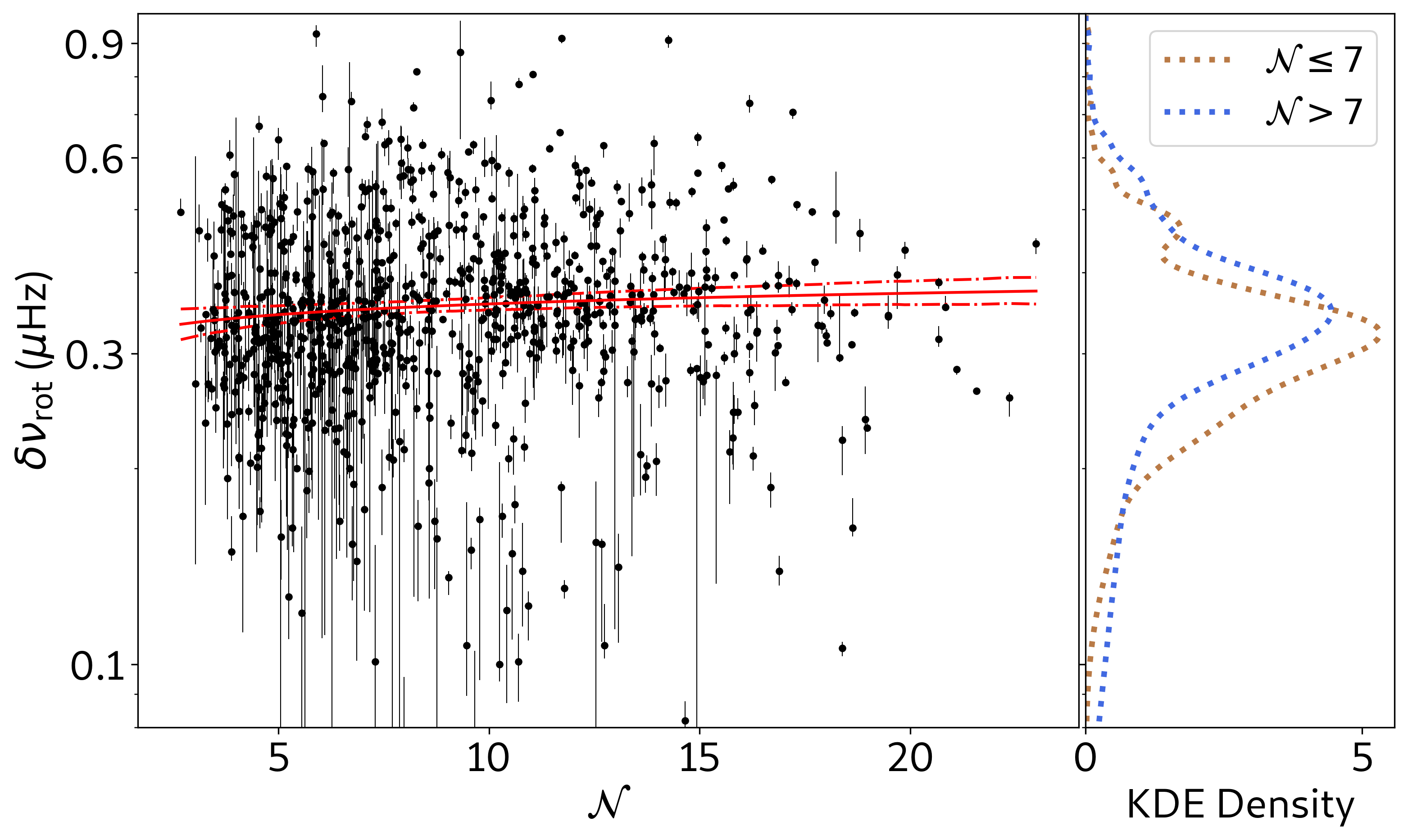}
    \caption{Observed mixed-mode rotational splitting $\delta\nu_{\mathrm{rot}}$ for the 877 \textit{Kepler} red giants in this study. The red line shows a fit of the form $\delta\nu_{\mathrm{rot}}= k\cdot\mathcal{N}^a$ to the data, with intervals indicating 95\% highest density regions of the posterior predictive distribution. The histograms visualize a slight shift in the mean observed value of $\delta\nu_{\mathrm{rot}}$ with evolution up the giant branch (increasing $\mathcal{N}$).}
    \label{fig:mean_rotation}
\end{figure}

The offset $\epsilon_g$ measures the phase offset of g-modes at the location of the convective boundary. Following \citet{Lindsay_2022} and \citet{Ong_2023}, our formulation of this offset, which is described in Appendix \ref{appendix:generation}, leads to values that are systematically smaller by $1/2$. This was intentional in the interest of consistency with the convention from previous literature. We clarify this in the plots and tables in this work by indicating that the presented offsets are in the form of $\epsilon_g-1/2$.

The upper turning point associated with the g-mode cavity for an early red giant is in the star's radiative region below the base of its convective zone. Analytic modelling by \citet{Pincon_2019} predicts that under the hypothesis of a thick evanescent region between p- and g- mode cavities (i.e., the weak coupling limit), the value of $\epsilon_g-1/2$ for early red giants remain constant at an asymptotic value of approximately 0.22. Figure \ref{fig:eg_hist} \MHupdate{shows that the $\epsilon_g-1/2$ distribution for stars with $\nu_{\mathrm{max}} > 110\mu$Hz peaks at this value, which provides observational support for the thick evanescent region hypothesis that is expected for early red giants in the weak mode coupling regime.
Stars with $\nu_{\mathrm{max}} \leq 110\,\mu$Hz have $\epsilon_g-1/2$ distributed differently because they begin experiencing a decrease in $\epsilon_g-1/2$ approaching the red giant luminosity bump (e.g., \citealt{Pincon_2019, Lindsay_2022}). To confirm this difference, we calculate a two-sided Kolmogorov-Smirnov (KS) test \citep{Massey_1951} for the distributions of $\epsilon_g-1/2$ for stars above and below $\nu_{\mathrm{max}} = 110\mu$Hz and find that the hypothesis that the two samples were drawn from the same distribution that can be rejected at a 1\% significance ($p\sim10^{-24}$). 
}
This provides observational support for the thick evanescent region hypothesis, which is expected given the weak mode coupling shown by these early red giants.

The point in evolution at which $\epsilon_g$ is observed to decrease ($\simeq 110\mu$Hz) is attributed to the g-mode cavity's upper turning point  
coinciding with the base of the convective zone \citep{Pincon_2019}. Models from \citet{Lindsay_2022} demonstrate that at this stage of evolution, $\epsilon_g$ progressively decreases and subsequently experiences rapid variations once reaching the red giant luminosity bump. Although our sample does not extend as far as the luminosity bump ($\nu_{\mathrm{max}}\sim50\mu$Hz, \citealt{Khan_2018}), we observe a gradual decrease in $\epsilon_g$ in Figure \ref{fig:eg_hist}. Given a fixed overshooting extent, models from \citet{Lindsay_2022} predict that the stage in evolution at which the  $\epsilon_g$ turnoff occurs is dependent on metallicity and mass. We observe no clear dependency of the location of this feature in $\epsilon_g-\nu_{\mathrm{max}}$ space across either metallicity or mass in Figure \ref{fig:eg_mass_met}. This would imply a non-consistent mixing length and overshooting prescriptions across stars in our sample, which is expected given that they are field stars.

\begin{deluxetable*}{ccccccc}[t]
\tablenum{2}
\tablecaption{Mixed-mode parameters for six metal-poor \textit{Kepler} red giants. \label{table:metal-poor}}
\tablewidth{0pt} 
\tablehead{ 
\colhead{KIC} & \colhead{$\Delta\Pi_1$} & \colhead{$q$} & \colhead{$\epsilon_g - 1/2$} & \colhead{$\delta\nu_{\mathrm{rot}}$} &  \colhead{[M/H]}  & \colhead{Source of reported metallicity} 
\\
\colhead{} & \colhead{(s)} & \colhead{} & \colhead{} & \colhead{($\mu$Hz)} &  \colhead{(dex)} &   \colhead{}
}
\startdata
3543446 & $78.56^{+0.04}_{-0.02}$ & $0.187^{+0.011}_{-0.012}$ & $0.202^{+0.012}_{-0.049}$ & $0.523^{+0.021}_{-0.019}$& $-1.23$& APOGEE \\
4671239 & $66.67^{+0.01}_{-0.06}$ & $0.291^{+0.040}_{-0.021}$ & $0.014^{+0.129}_{-0.010}$ & $0.815^{+0.017}_{-0.007}$& $-2.44$& \citet{Arthur_2022}\\
7948268 & $76.69^{+0.06}_{-0.02}$ & $0.194^{+0.009}_{-0.007}$ & $0.180^{+0.033}_{-0.066}$ & $0.272^{+0.008}_{-0.009}$& $-1.28$& APOGEE\\
8144907 & $84.83^{+0.03}_{-0.02}$ & $0.309^{+0.008}_{-0.007}$ & $0.205^{+0.015}_{-0.024}$ & $0.335^{+0.006}_{-0.020}$& $-2.70$& This study\\
11460115 & $74.57^{+0.01}_{-0.02}$ & $0.201^{+0.016}_{-0.008}$ & $0.152^{+0.030}_{-0.006}$ & $0.582^{+0.005}_{-0.009}$& $-1.30$& This study\\
11704816 & $70.96^{+0.05}_{-0.03}$ & $0.198^{+0.024}_{-0.046}$ & $0.081^{+0.026}_{-0.128}$ & $0$& $-1.75$& This study\\
\enddata
\end{deluxetable*}

\subsection{Distribution of Observed Mixed-mode Rotational Splittings}\label{section:drot}
A total of 882 out of 1,068 stars from our sample have detectable mixed-mode rotational splittings, with their distribution of $\delta\nu_{\mathrm{rot}}$ as a function of mixed-mode density $\mathcal{N} = \Delta\nu/(\Delta\Pi_1\cdot\nu_{\mathrm{max}}^2)$ shown in Figure \ref{fig:summary}d. Following \citet{Gehan_2018}, we fit a relation of the form $\delta\nu_{\mathrm{rot}}\propto\mathcal{N}^a$. Our fit over the full sample in Figure \ref{fig:mean_rotation} yields an exponent $\langle a \rangle =0.05\pm0.02$, indicating a gradual increase in rotational frequency splittings as the star evolves up the giant branch. The histograms in Figure \ref{fig:mean_rotation} similarly illustrate this observation, whereby $\delta\nu_{\mathrm{rot}}$ for giants with $\mathcal{N}\leq7$ peaks at a value smaller than those with $\mathcal{N}>7$ by approximately $0.03\mu$Hz. \MHupdate{ To quantify this observation, we first conduct a two-sided KS test for $\delta\nu_{\mathrm{rot}}$ values between the sample of giants with $\mathcal{N}\leq7$ and those with $\mathcal{N}>7$. We compute $p\sim10^{-6}$, such that we can reject the null hypothesis with at least a 1\% significance. Next, we perform a Bayesian fit to estimate the average $\delta\nu_{\mathrm{rot}}$ of each sample by assuming a Student-T likelihood and a $\mathcal{N}(0.2, 1)$ prior. we estimate the values of $\delta\nu_{\mathrm{rot}}$ for the $\mathcal{N}\leq7$ sample to be centered at 0.33 $\mu$Hz as opposed to 0.36 $\mu$Hz for the $\mathcal{N}>7$ sample. Although we find that the fitted posterior distribution of the average $\delta\nu_{\mathrm{rot}}$ for the $\mathcal{N}\leq7$ sample is credibly different compared to the $\mathcal{N}>7$ sample, they are not highly statistically significant given the typical uncertainty of our $\delta\nu_{\mathrm{rot}}$ measurements. This interpretation is consistent with our measured $\langle a \rangle$, which does not significantly depart from the the value measured by \citet{Gehan_2018} of $\langle a \rangle =0.01\pm0.03$.}


In addition to a global fit to all stars with detected rotational splittings, we bin our sample based on either their mass or metallicity, and perform hierarchical fits of the form $\delta\nu_{\mathrm{rot}}\propto\mathcal{N}^a$ across each bin. The fitted coefficients, which are tabulated in Table \ref{table:core_rot}, show that the fitted exponents across bins are consistent with zero. \MHupdate{Therefore, our results do not identify statistically significant correlations between $\delta\nu_{\mathrm{rot}}$ with mass and metallicity, in agreement with the results from \citet{Gehan_2018}.}

\begin{table}[ht]
\tablenum{3}
\centering
\caption{Parameter values of the fits of the form $\delta\nu_{\mathrm{rot}}\propto\mathcal{N}^a$ to the mixed-mode rotational splittings. The full sample of stars with detectable mixed-mode rotational splittings comprises 877 stars, while the subset for which the binning in mass and metallicity is applied comprises 570 stars. }\label{table:core_rot}
\begin{tabular}{cc}
\toprule
\multicolumn{2}{c}{\textbf{Full Sample}} \\
\midrule
\multicolumn{2}{c}{$\langle a \rangle=0.05\pm0.02$}\\
\midrule
\multicolumn{2}{c}{\textbf{Binned in Mass, $M$}} \\ 
\midrule
$M / M_{\odot}< 1.00$ & $a=0.03\pm0.03$ \\
1.00 $\leq M/ M_{\odot} < $  1.15 & $a=0.02\pm0.03$ \\
1.15 $\leq$ $M/ M_{\odot} < $ 1.30 & $a=0.02\pm0.03$\\
1.30 $\leq$ $M/ M_{\odot}  < $ 1.45 & $a=0.05\pm0.03$\\
1.45 $\leq$ $M/ M_{\odot} < $ 1.60 & $a=0.02\pm0.03$\\
$M/ M_{\odot}$ $\geq$ 1.60 & $a=0.01\pm0.04$ \\
\midrule
\multicolumn{2}{c}{\textbf{Binned in Metallicity, $\mathrm{[M/H]}$ (dex)}} \\
\midrule
$\mathrm{[M/H]}$ $<$ -0.5  & $a=0.03\pm0.03$\\
-0.5 $\leq$ $\mathrm{[M/H]}$ $<$ -0.3 & $a=0.03\pm0.03$\\
-0.3 $\leq$ $\mathrm{[M/H]}$ $<$ -0.1 & $a=0.02\pm0.03$\\
-0.1 $\leq$ $\mathrm{[M/H]}$ $<$ -0.1 & $a=0.04\pm0.03$\\
+0.1 $\leq$ $\mathrm{[M/H]}$ $<$ +0.3 & $a=0.04\pm0.03$\\
$\mathrm{[M/H]}$ $\geq$ +0.3 & $a=0.03\pm0.03$\\
\bottomrule
\end{tabular}
\end{table}

\section{Limitations of this Study}\label{section:limitations}
The precision of our measurements in this study is limited by several simplifying assumptions in our modelling of the mixed-modes. 
\subsection{Coupling Factor}
We assume that the coupling factor $q$ remains constant, although coupling strength variations over the observable frequency range of dipole mixed-modes are expected in theory \citep{Shibahasi_1979, Pincon_2020} and from stellar models \citep{Jiang_2014, Cunha_2015, Mosser_2017, Hekker_2018, Jiang_2020}. The lack of a frequency dependence in our formalism of the coupling factor in this work can lead to systematic offsets in our inference of $q$, particularly for the more evolved stars in our sample with thicker evanescent regions \citep{Pincon_2020}. Such offsets may explain the observed scatter in $q$ for stars with $\Delta\nu\apprle 9\mu$Hz  (as seen in Figures \ref{fig:summary}b, \ref{fig:dnu_q}, and \ref{fig:dnu_q_mass}), which are not predicted by evolutionary models \citep{Jiang_2020}. In further work, we will aim to incorporate linear dependencies of $q$ with frequency \textemdash{} such as those proposed by \citet{Cunha_2015} and \citet{Jiang_2020} \textemdash{} into the framework of the Bayesian optimization as an additional free parameter. Additionally, in our construction of $\zeta$ we adopt
an empirically motivated scaled form of $d_{01}$ for Equation \ref{eqn:asymptotic}. Including $d_{01}$ as a free parameter to more accurately model $\nu_p$ will further improve the precision in our measurements of $q$. 
Finally, we point out that observational values of $q$ are known to be systematically larger compared to those from stellar models \citep{Ong_2023}. Given the intrinsic nature of this systematic offset across low-luminosity giants, we do not expect it to affect our phenomenological conclusion regarding the identification of metal-poor stars using $q$.

\subsection{Solving for Mixed-mode Frequencies}
As described in Appendix \ref{appendix:zeta}, we use the quick and simple approach of searching for a candidate frequency corresponding to a zero crossing in Equation \ref{eqn:thetaP_thetaG} across a grid of frequencies. This may lead to imprecise root solving of Equation \ref{eqn:thetaP_thetaG}, resulting in the observed discontinuities near the pure dipole mode frequencies in the generated stretched \'echelle diagrams. 
The systematic uncertainties are expected to contribute to part of our reported uncertainties. With an increased computational budget in future, the use of more sophisticated root solvers will improve the precision of our inference. 

\subsection{Observed Mixed-mode Rotational Splitting}
\begin{figure}
    \centering
    \includegraphics[width=1.\columnwidth]{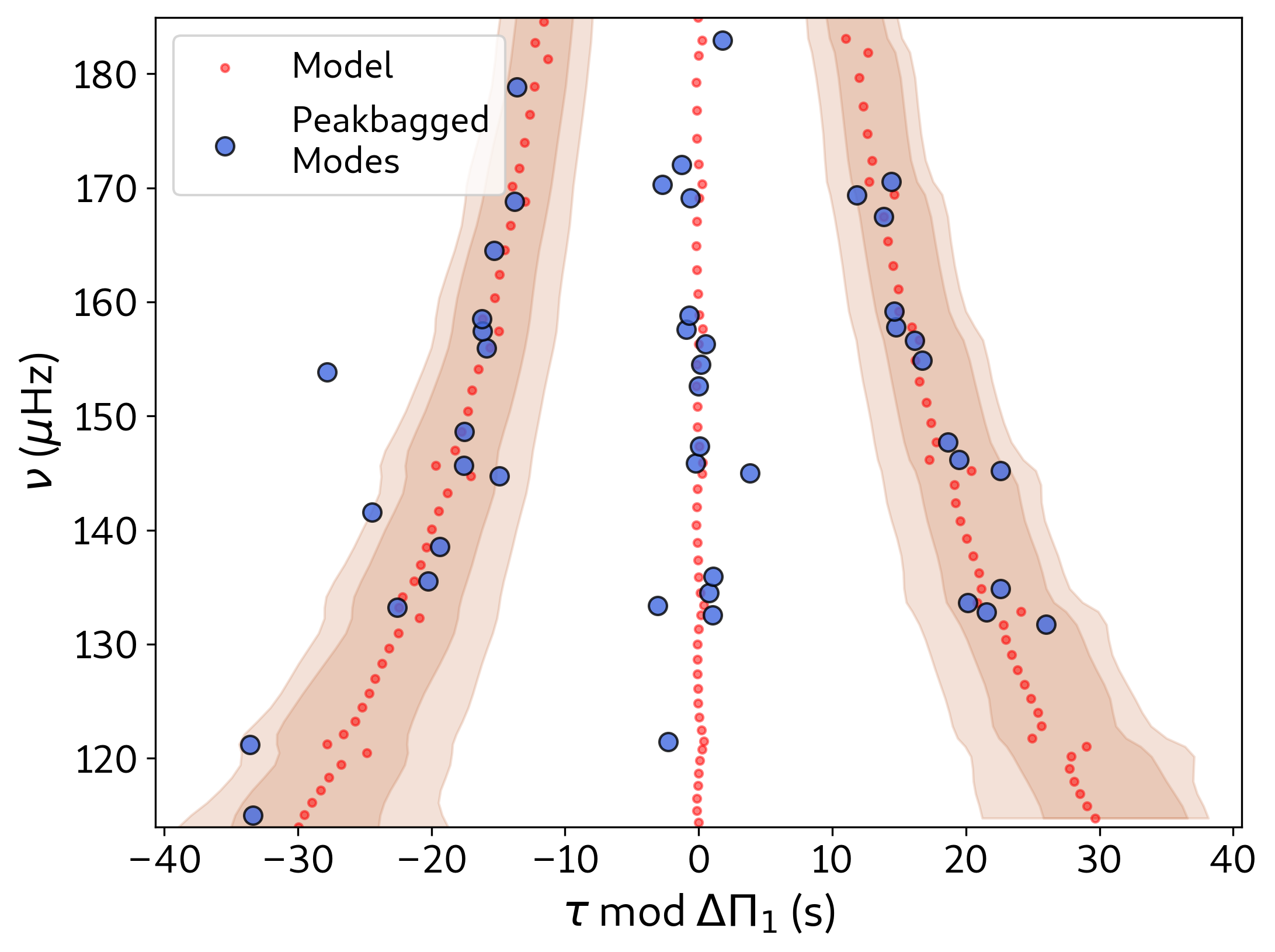}
    \caption{Stretched period \'echelle diagram of KIC 3973247. The contours represent 95\% and 99\% highest density intervals of the BayesOpt samples in Figure \ref{fig:clustering}, and indicate the uncertainty of the fit between the model and observed mixed-modes. }
    \label{fig:echelle_uncertainty}
\end{figure}

Our work adopts a simplistic treatment of $\delta\nu_{\mathrm{rot}}$. This measurement corresponds to the rotational splitting of the g-mode dominated mixed-modes, which have contributions from both the core and the envelope of red giants. The computation of $\delta\nu_{\mathrm{rot}}$ in our framework assumes slow surface rotation and subsequently a negligible contribution to the rotationally splittings from the stellar envelope \citep{Goupil_2013}. Following \citet{Gehan_2018}, we suggest caution in interpreting these measurements directly as a proxy for mean core rotation rates, particularly for stars lower on the giant branch where the envelope contribution cannot be neglected \citep{Goupil_2013}. Furthermore, our computed value of $\zeta$ for each rotationally split mixed-mode adopts the value corresponding to that from non-rotationally split mode (i.e., $\zeta(\nu_{n,m}) = \zeta(\nu_{n, 0})$). As discussed by \citet{Mosser_2017}, estimating $\zeta$ at the correct frequencies for rotational multiplets allows the detection of asymmetrical rotational splittings, some of which are hypothesized to be caused by strong core magnetic fields \citep{Li_2022}. Our method is thus expected to be insensitive to relatively small splitting asymmetries as their contributions would be further diminished by our construction of $\zeta$ for the rotationally split mixed-modes.
Despite the approximations taken in our study, our reported estimates and their corresponding uncertainties (Figure \ref{fig:echelle_uncertainty}) provide strong constraints on the range of ($\Delta\Pi_1, q, \epsilon_g, \delta\nu_{\mathrm{rot}}$) for each low luminosity red giant in this study. These constraints may form useful priors for more detailed fits using more sophisticated mixed-mode pattern prescriptions.

\subsection{Assumption of Three Rotationally-Split Mixed-mode Components}\label{section:triplet_assump}
\MHupdate{
Out of simplicity, we perform the forward modelling BayesOpt step in Section \ref{section:forward_modelling} assuming that the peakbagged modes contain $m=0, \pm 1$ mixed-mode rotational components. This is, however, not all the case because stars viewed equator-on may only have $m=\pm1$ modes visible whereas those viewed pole-on will only show $m=0$ modes. Without any prior knowledge about the inclination angle of the star, our assumption naturally leads to errors in our fitting of mixed-mode parameters. To mitigate these errors once having a solution from the BayesOpt fit, we determine the fractional contribution of each rotational component to that fit. This is done by removing synthetic modes of one particular $m$ at a time and recalculating the corresponding BayesOpt loss (Equation \ref{eqn:freq-tau_matching}). By analyzing the fractional
contribution of different rotational components to the BayesOpt loss, we can identify which the correct number of modes to fit. We detail this additional analysis in Appendix \ref{appendix:triplet}.
For instance, we expect a star viewed pole-on to have all of its contribution to the BayesOpt loss contained within the $m=0$ modes. However, if our analysis reveals that the BayesOpt loss is concentrated only in the matching of $m=+1$ modes, then we can identify this scenario as an incorrectly singlet and forward modelling is rerun assuming only a single rotational component.}

\MHupdate{
We additionally note that the BayesOpt analysis in this work does not use any information about inclination angles from the peakbagging, but only uses information about the mode frequencies. Therefore, our measured $\delta\nu_{\mathrm{rot}}$ and more generally the
mixed-mode parameters estimated in this study are directly dependent upon the quality of the peakbagging. For this reason \textemdash{} as well as that of reproduciblity \textemdash{} we have included all peakbagged modes in our repository.
}

\subsection{Additional Curvature}
\begin{figure}
    \centering
    \includegraphics[width=1.\columnwidth]{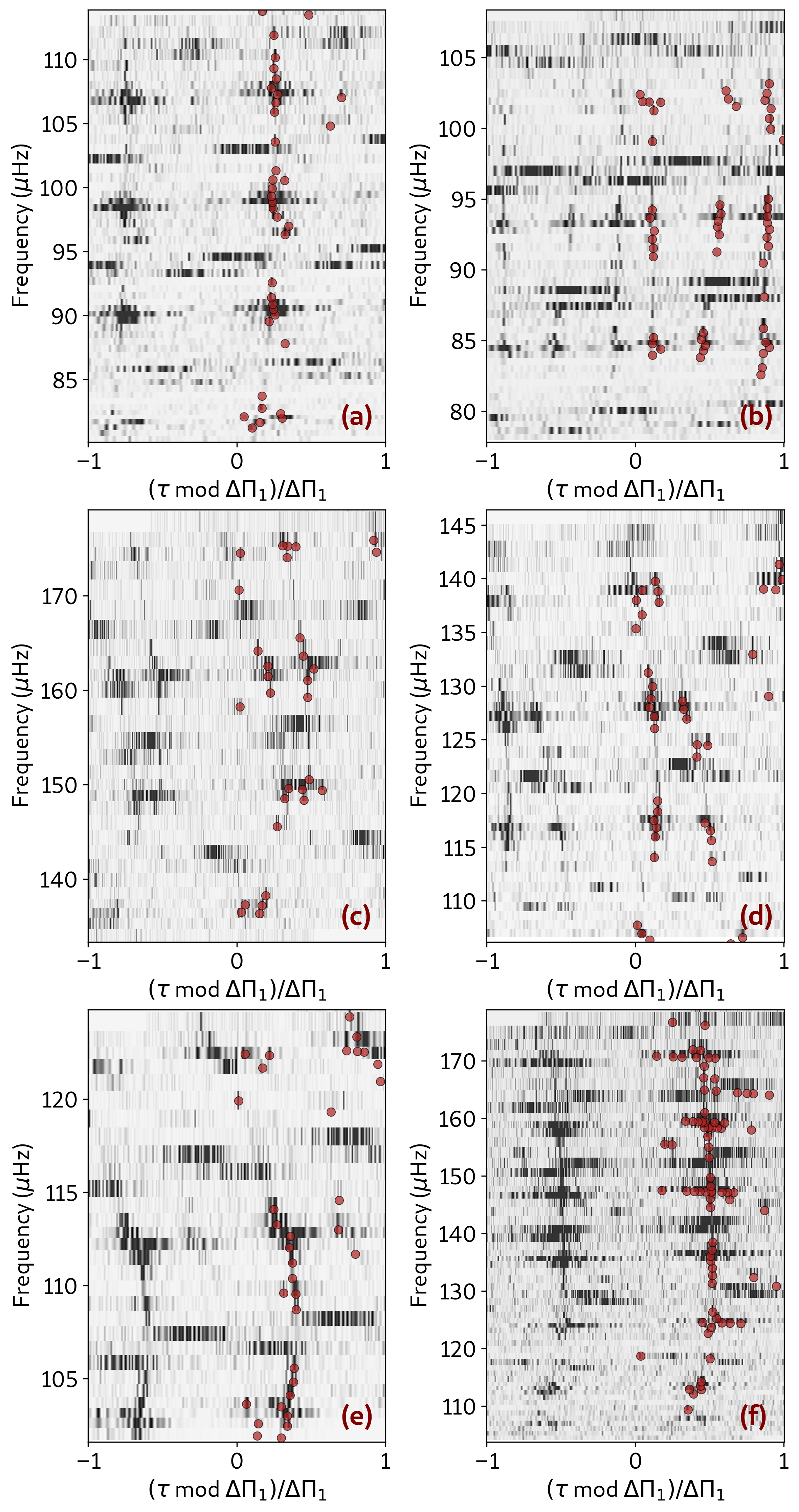}
    \caption{Replicated stretched period-power \'echelle diagrams for six stars from the \citet{Gehan_2018} sample that show curvature in their g-mode pattern. These are (\textit{a}) KIC 6786177, (\textit{b}) KIC 6936091, (\textit{c}) KIC 6953912, (\textit{d}) KIC 7458743, (\textit{e}) KIC 10973397, and (\textit{f}) KIC 11087371. Also shown are peakbagged dipole mixed-mode frequencies to guide the eye. The mixed-mode parameters to generate these diagrams are listed in Table \ref{table:curvature}. The construction of this diagram follows the implementation presented in \citet{Ong_2023}.}
    \label{fig:curvature}
\end{figure}

Our method does not account for the presence of additional curvature in the stretched period \'echelle diagram that are hypothesized to be due to glitches in the star's Brunt-V\"{a}is\"{a}l\"{a} frequency profile (e.g., \citealt{Mosser_2015, Lindsay_2022, Vrard_2022}), or by interior magnetic fields \citep{Li_2022, Deheuvels_2023}. The presence of such a variation will result in poor fits during the optimization. We identify 6 \textit{Kepler} giants from the \citet{Gehan_2018} sample with previously unidentified curvature in Figure \ref{fig:curvature}. Due to their anomalous nature, we do not include these stars in Table \ref{table:mixed_mode_parameters}, but in a separate Table \ref{table:curvature} that includes the $(\Delta\Pi_1, q)$ used to generate the plots in Figure \ref{fig:curvature}. We note that KIC 6936091 (Figure \ref{fig:curvature}b) shows both curvature and rotational splitting asymmetry.

\begin{deluxetable}{ccc}[t]
\tablenum{4}
\tablecaption{Mixed-mode parameters used to generate the replicated stretched period-power e\'chelle diagrams in Figure \ref{fig:curvature}. \label{table:curvature}}
\tablewidth{0pt} 
\tablehead{ 
\colhead{KIC} & \colhead{$\Delta\Pi_1$} & \colhead{$q$}
\\
\colhead{} & \colhead{(s)} & \colhead{}
}
\startdata
6786177 & 71.91 & 0.110 \\
6936091 & 75.25 & 0.118 \\ 
6953912 & 76.09 & 0.140 \\
7458743 & 76.15 & 0.136 \\
10973397 & 69.47 & 0.106 \\
11087371 & 77.26 & 0.120 \\
\enddata
\end{deluxetable}





\section{Conclusion}
Our work can be summarized as follows:

\begin{enumerate}
    \item We determined mixed-mode parameters in the form of the asymptotic g- mode period spacing ($\Delta\Pi_1$), mode coupling factor ($q$), g-mode phase offset ($\epsilon_g$), and observed mixed-mode rotational splitting ($\delta\nu_{\mathrm{rot}}$) for 1,074 low-luminosity red giants. To achieve this, we have peakbagged mode frequencies of 1,146 stars, which includes 1,140 that belong to the  \citet{Gehan_2018} sample. 

    \item We presented a framework for estimating mixed-mode parameters ($\Delta\Pi_1, q, \epsilon_g, \delta\nu_{\mathrm{rot}}$) from peakbagged results by incorporating Bayesian optimization firstly in the power spectrum of the stretched period spectrum, and then in a forward modelling step where model frequencies are matched to peakbagged frequencies. Our peakbagging results available at \dataset[10.5281/zenodo.7888633]{\doi{10.5281/zenodo.7888633}}, while the code to generate model frequencies, and fits to the mixed-mode pattern are available at \url{https://github.com/jsk389/BOChaMM/}.

    \item We investigated correlations of the mixed-mode parameters with mass and metallicity for 684 red giants with spectroscopic parameters from APOGEE DR17. In our investigation of $\Delta\Pi_1$, we identify a similar group to that found by \citet{Deheuvels_2022} in the form of massive ($M>1.4M_{\odot}$) red giants that are positioned below the `degenerate sequence'. 

    \item We confirmed modelling predictions of a correlation between $q$ and the star's bulk metallicity [M/H]. Low metallicity stars have fitted values of $q$ systematically offset to higher values, with very low-metallicity stars ([M/H] $<$ -2 dex) achieving coupling factors as large as $\sim0.3$. A correlation between $q$ and stellar mass appears present, but its effect is observed to be significantly weaker than that by metallicity.

    \item We presented the evolution of $\epsilon_g$ up the lower giant branch. The convergence of $\epsilon_g-0.5$ at 0.22 for stars lower on the giant branch, and the departure of $\epsilon_g$ to lower values when approaching the luminosity bump agrees with modelling results from \citet{Pincon_2019} and supports early observational trends presented by \citet{Mosser_2018}. We found no distinct correlation between the correlation of $\epsilon_g$ with mass or metallicity.

    \item We provided independent confirmation of the \citet{Gehan_2018} result that no statistically significant correlation between $\delta\nu_{\mathrm{rot}}$ and mass or metallicity can be identified.

    \item We identified 6 new red giants showing anomalous variations in their mixed-mode spectrum indicative of buoyancy glitches or core magnetic fields.
\end{enumerate}

Altogether, our observational study reveals how the interiors of red giants as probed by their mixed-modes are distributed at a population level, which will
add to the utility of asteroseismology in studying stellar evolution across large surveys like TESS \citep{Ricker_2015} and PLATO \citep{Rauer_2014}.


\begin{acknowledgments}
Funding for the TESS mission is provided by the NASA Explorer Program. M.H. acknowledges support from NASA through the NASA Hubble
Fellowship grant HST-HF2-51459.001 awarded by the Space Telescope
Science Institute, which is operated by the Association of Universities for Research in Astronomy, Incorporated, under NASA contract NAS5-26555. D.H. acknowledges support from the Alfred P. Sloan Foundation, the National Aeronautics and Space Administration (80NSSC21K0652, 80NSSC20K0593), and the Australian Research Council (FT200100871). We express warm gratitude to Christopher Lindsay and Joel Ong for fruitful discussions and for the implementation of the stretched period-power \'echelle diagrams.
\end{acknowledgments}

\appendix

\section{Trust Region Bayesian Optimization} \label{appendix:turbo}

\begin{figure*}
    \centering
	\includegraphics[width=1.\columnwidth]{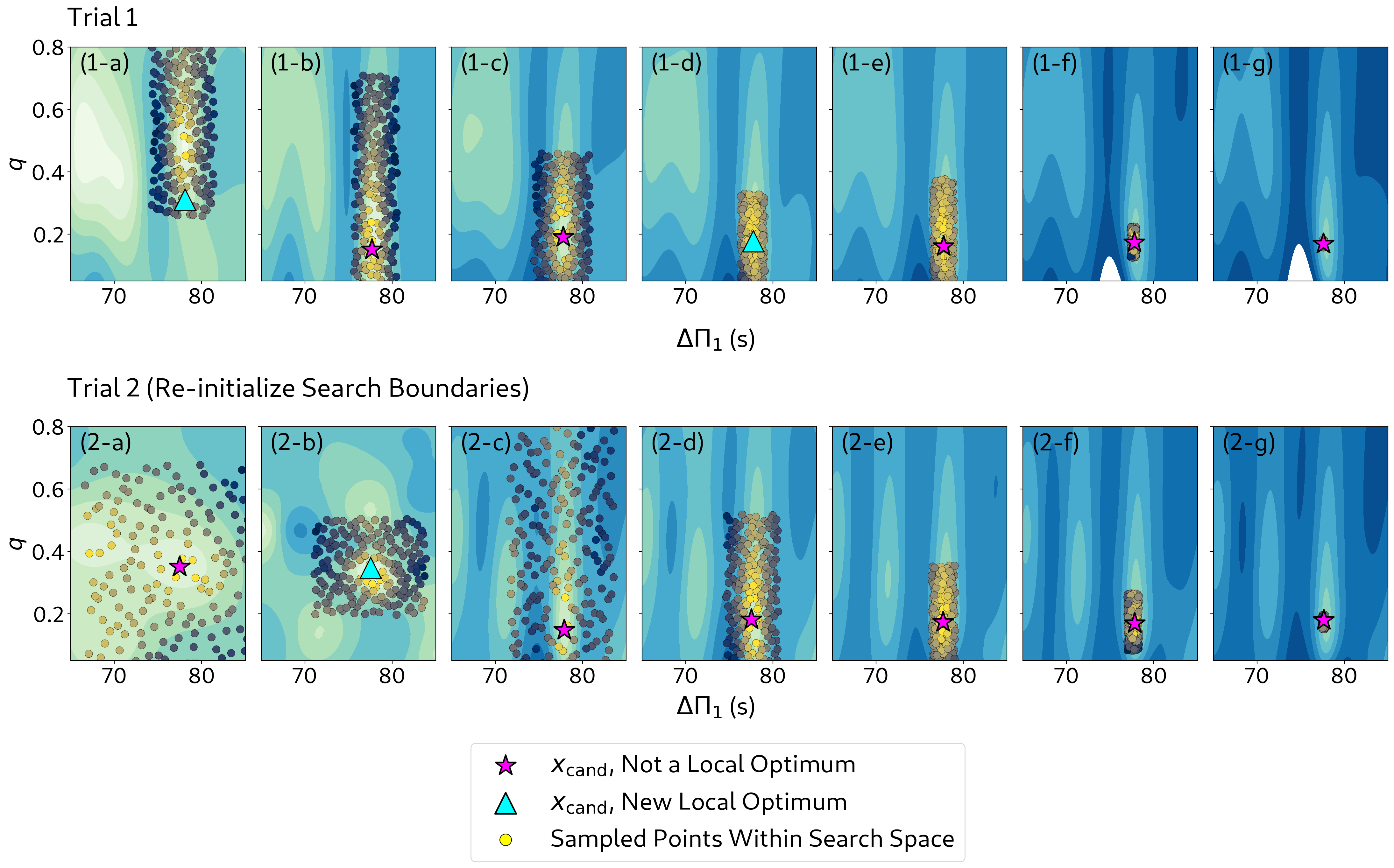}
    \caption{Visualization of trust region Bayesian optimization (\texttt{TuRBO}) over a 2D space. The goal is to find the optimal ($\Delta\Pi_1$, $q$) that maximizes the objective function, which here is the power at $\Delta\Pi_1$
    within the power spectrum of the stretched period oscillation spectrum (see Section). Sequential iterations of \texttt{TuRBO} are shown from left to right.
    In each iteration, a region of ($\Delta\Pi_1$, $q$) 
    is uniformly sampled (dots), \MHupdate{which forms the input to the Gaussian process component in \texttt{TuRBO} that provides a smooth and continuous approximation to the true parameter space}. Here, lighter colors correspond to a larger power estimated at corresponding $\Delta\Pi_1$. The contours show the approximation of the Gaussian process to the entire 2D space, including regions outside the sampled space. As more iterations progress, the contour approximations become more detailed and accurate in modelling the whole 2D space. In each sampled region, $x_{\mathrm{cand}}$ is determined as the optimal value from the \MHupdate{approximation by the Gaussian process. Because the approximation is smooth and continuous, $x_{\mathrm{cand}}$ is not restricted to follow any specific discretization of the parameter space. } If $x_{\mathrm{cand}}$ is \MHupdate{the best solution across all current iterations} (blue triangle), the search region in the next iteration is expanded about $x_{\mathrm{cand}}$. Otherwise, the search region is shrunk in the following iteration. Once the search region shrinks beyond a certain limit from consecutive failures in identifying a new global optimum, the search is re-initialized \textemdash{} here shown as Trial 2. }
    \label{fig:turbo}
\end{figure*}

In this work, we use a variant of Bayesian optimization known as Trust Region Bayesian Optimization \citep[\texttt{TuRBO}]{Eriksson_2019}. This variant has the same characteristics of BayesOpt algorithms described in Section \ref{section:bayesopt} with several modifications. First, the algorithm only identifies $x_{\mathrm{cand}}$ from points within a local `trust' region of the parameter space in one iteration. The spatial extent of this region is altered in subsequent iterations depending on whether or not a new local optimum can currently be found from a candidate within this region. If it is found, the extent of the search region in the next iteration is expanded about $x_{\mathrm{cand}}$, otherwise it is shrunk. Therefore, successive failures will continuously shrink the search region until it reaches below a certain user-defined tolerance, at which the region re-initializes to a different location in parameter space. A visualization of this process is shown in Figure \ref{fig:turbo}. Another important modification in \texttt{TuRBO} is its acqusition function, which uses a search strategy based on Thompson sampling \citep{Thompson_1933}. In particular, the selection of $x_{\mathrm{cand}}$ is directly guided by the GP, in the sense that $g(x_{\mathrm{cand}})$ is required to be the optimum value across the search region as predicted by the GP. In our work, we use \texttt{TuRBO} over other BayesOpt algorithms because we find its adaptive search region strategy to be highly efficient in exploring the mixed-mode parameter space in as few iterations as possible while  avoiding excessive exploration within local optima. Additionally, the algorithm's internal use of \texttt{GPyTorch} \citep{Gardner_2018} offers GPU support that significantly speeds up runtime over the large sample of stars investigated in this work. The code for the algorithm is available at \url{https://github.com/uber-research/TuRBO}.

\section{Detailed Implementation}

\subsection{Construction of Mixed-mode Stretching Factor} \label{appendix:zeta}
The stretching factor $\zeta(\nu)$ measures how strongly the mixed-mode behaves as a g-mode and is defined as  the ratio of mode inertia in the g cavity over the total mode inertia \citep{Deheuvels_2012,Goupil_2013}. To construct this function, we consider each dipole pure pressure mode, $\nu_{n_p}$ within the frequency range of the power spectrum containing oscillations, for which we estimate $\Delta\nu_p$ as the frequency difference between consecutive radial orders (i.e., $\nu_{n_p+1} - \nu_{n_p}$). Values of $\nu_{n_p}$ are estimated using the asymptotic form:
\begin{equation}\label{eqn:asymptotic}
    \frac{\nu_{n_p}}{\Delta\nu} = n_p + \frac{l}{2} + \epsilon - d_{01} + \frac{\alpha}{2}\bigg(n_p - \frac{\nu_{\mathrm{max}}}{\Delta\nu} \bigg)^2 , 
\end{equation}
where $\Delta\nu$ is the `average' large frequency spacing across the oscillation spectrum, and $d_{01} = -0.056 - 0.002 \log \Delta\nu$, which is based on the fits to the `universal pattern' of red giant oscillations along the red giant branch \citep{Mosser_2011}.

Given the asymptotic dipole mode period spacing, $\Delta\Pi_1$ and its corresponding mode coupling factor $q$, we define the following:
\begin{equation}
    \theta_p = \frac{\pi(\nu-\nu_p)}{\Delta\nu_p} \quad\text{and}\quad \mathcal{N} = \frac{\Delta\nu}{\Delta\Pi_1\,\nu_p^2},
\end{equation}
where $\mathcal{N}$ is also known as the mixed-mode density. With these quantities, we compute the following 

\begin{equation}
    f = 1 + \frac{q/\mathcal{N}}{q^2\cos(\theta_p)^2 + \sin(\theta_p)^2} \quad\text{and}\quad \zeta_{\mathrm{max}} = \frac{1}{1+q/\mathcal{N}}.
\end{equation}
The model for $\zeta$ per radial order is thus given by
\begin{equation}
    \hat{\zeta}(\nu_p) = \frac{1}{f} + (1-\zeta_{\mathrm{max}}).
\end{equation}
These are combined across all relevant radial orders to construct $\hat{\zeta}(\nu)$. We account for deviations of $\zeta_{\mathrm{max}}$ from unity by a linear interpolation ($\mathcal{I}$) of $\zeta_{\mathrm{max}}$ across $\nu$ such that our final model of $\zeta$ is given by:
\begin{equation}
   \zeta(\nu)  = \hat{\zeta}(\nu) - (1-\mathcal{I}).
\end{equation}

\subsection{Generation of Mixed-mode Frequencies}\label{appendix:generation}
\noindent For each $\nu_p$, we define 
\begin{equation}
   n_{\mathrm{min}}  = \left \lfloor{\frac{1}{\Delta\Pi_1  \nu_{0, +}} - 1/2 - \epsilon_g}\right \rfloor \quad\text{and}\quad n_{\mathrm{max}}  = \left \lceil{\frac{1}{\Delta\Pi_1  \nu_{0, -}} - 1/2 - \epsilon_g}\right \rceil,
\end{equation}
where $ \nu_{0, -}$ and $ \nu_{0, +}$ are frequencies of the radial modes directly below and above $\nu_p$, respectively. These determine the range of integer values $|n_g|\in[n_{\mathrm{min}}, n_{\mathrm{max}}]$, from which we have (e.g., Equation 56 in \citet{Hekker_2017}): 
\begin{equation}
    \nu_g = \frac{1}{\Delta\Pi_1(|n_g| + 1/2 + \epsilon_g)},
\end{equation}
which represents the uncoupled g-mode frequencies in the vicinity of the radial order containing $\nu_p$. Here, we define the lower and upper bounds of a grid containing 1,000 trial values of $\nu$, where $\nu\in[\nu_{g, -}, \nu_{g, +}]$ and
\begin{equation}
    \nu_{g, -} = \frac{1}{\Delta\Pi_1(|n_g| + 1/2 + \epsilon_g + 1/2)} \quad\text{and}\quad \nu_{g, +} = \frac{1}{\Delta\Pi_1(|n_g| + 1/2 + \epsilon_g - 1/2)}
\end{equation}
respectively. By iterating over each value in the $\nu_g$ based on $|n_g|\in[n_{\mathrm{min}}, n_{\mathrm{max}}]$, we compute
\begin{equation}
    \theta_g = \frac{\pi}{\Delta\Pi_1}\bigg(\frac{1}{\nu} - \frac{1}{\nu_g}\bigg) + \frac{\pi}{2},
\end{equation}
and search for values within the grid of $\nu$ that corresponds to zero crossings in the equation
\begin{equation}\label{eqn:thetaP_thetaG}
    \tan\theta_p = q\tan\theta_g .
\end{equation}
In other words, we identify the frequencies at which a change of sign occurs for the function $\tan\theta_p - q\tan\theta_g$.
The solutions found across these iterations provide the mixed-mode frequencies, $\nu_m$, associated with a particular $\nu_{n_p}$. 

An input power spectral density of an oscillating star contains a grid of frequency values ($\nu$), and the power associated with each frequency bin. With the construction of $\zeta$ as detailed in Appendix \ref{appendix:zeta}, $d\tau$ (as defined in Equation \ref{eqn:tau}) is constructed over $\nu$, and $\tau(\nu)$ is subsequently determined using the integral (or cumulative sum) over the $d\tau$ values. For each computation of $\tau$, 
we determine the offset of $\tau\,$mod$\,\Delta\Pi_1$ from zero, $\epsilon_{\mathrm{shift}}$, which allows us to center the $m=0$ ridge in the stretched period \'echelle diagram at $\tau\,$mod$\,\Delta\Pi_1 = 0$.

Next, given a series of input mode frequencies, $\nu_{s}$, their corresponding stretched periods, $\tau_{s}$, are evaluated by the linear interpolation of $\tau(\nu)$ at $\nu=\nu_{s}$. Similarly, values of $\zeta_{s}$ are determined from the interpolation of the existing $\zeta$ function at $\nu=\nu_{s}$.

In the presence of core rotation, the rotational splitting for a simulated mixed-mode is computed as $\delta\nu_{\mathrm{rot}} = \zeta_(\nu_{n,0})\cdot\delta\nu_{\mathrm{split}}$ \citep{Goupil_2013}, where $\zeta(\nu_{n,0})$ is the evaluation of $\zeta$ for a mixed-mode frequency in the absence of rotation \citep{Mosser_2018} and $\delta\nu_{\mathrm{split}} \simeq \frac{1}{2} \langle\frac{\Omega}{2\pi}\rangle_{\mathrm{core}}$ \citep{Goupil_2013, Mosser_2015,Gehan_2018}. Values of $\tau$ for the rotationally-split mixed-modes are computed as $\tau(\nu_{\mathrm{n,0}} \pm \delta\nu_{\mathrm{rot}})$, which is then offset with the previously computed $\epsilon_{\mathrm{shift}}$ to correctly center the location of the rotationally-split ridges in the stretched period \'echelle diagram. 

\section{Stretched Period \'Echelle Diagrams of Metal-Poor \textit{Kepler} Giants}\label{appendix:metal-poor_ech}
\begin{figure*}
    \centering
	\includegraphics[width=1.\columnwidth]{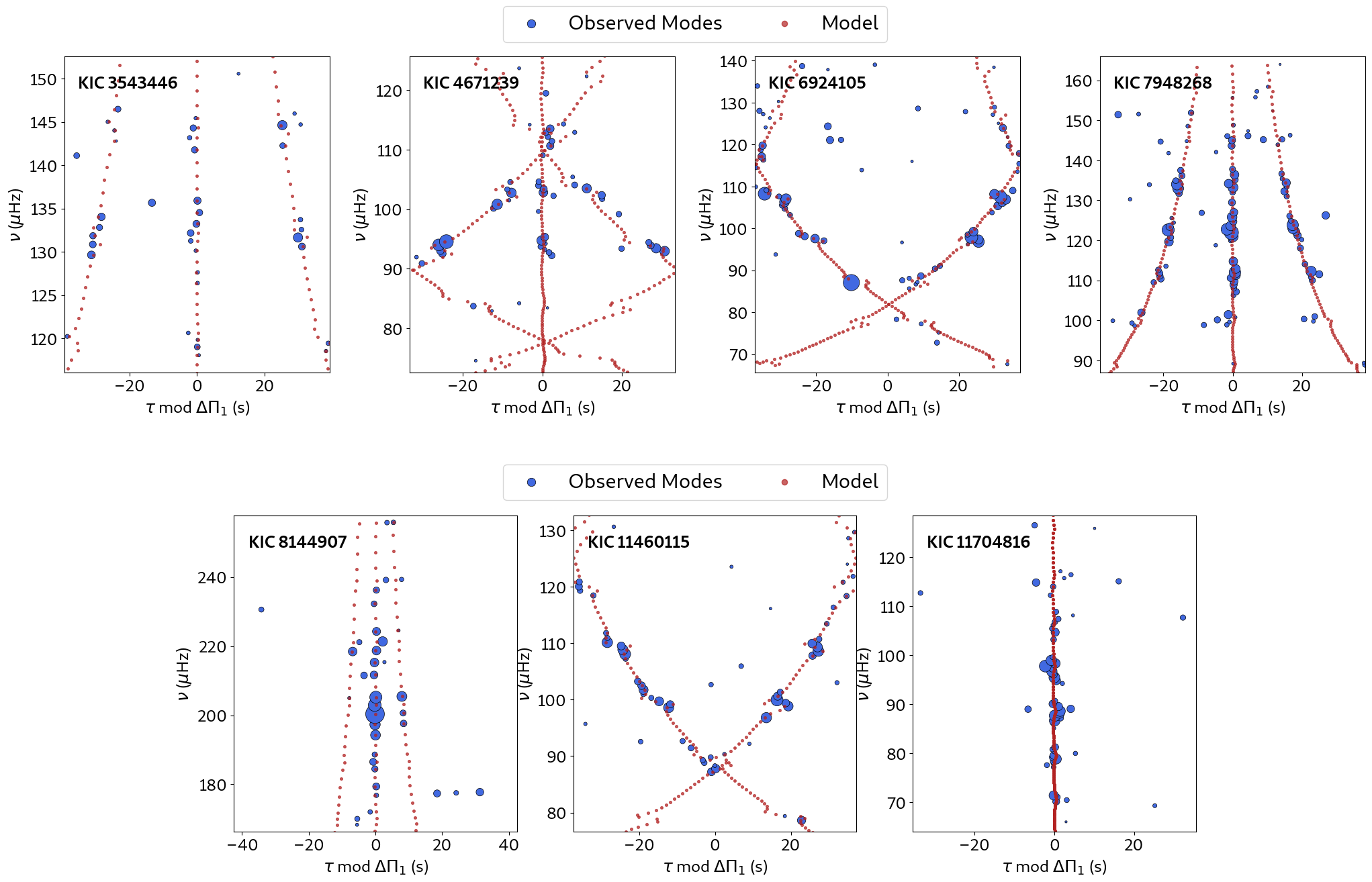}
    \caption{Stretched period \'echelle diagrams for the seven metal-poor \textit{Kepler} low-luminosity red giants. } 
    \label{fig:metal-poor_ech}
\end{figure*}

Figure \ref{fig:metal-poor_ech} shows the fitting of the mixed-mode pattern of the seven metal-poor \textit{Kepler} giants in Section \ref{section:q}. The values of ($\Delta\Pi_1, q, \epsilon_g$, and $\delta\nu_{\mathrm{rot}}$) and their metallicities are tabulated in Table \ref{table:metal-poor}. 

\section{Mass and Metallicity Correlation of Stars in our Sample}
\label{appendix:mass_met_corr}
\begin{figure}
    \centering
    \includegraphics[width=0.5\columnwidth]{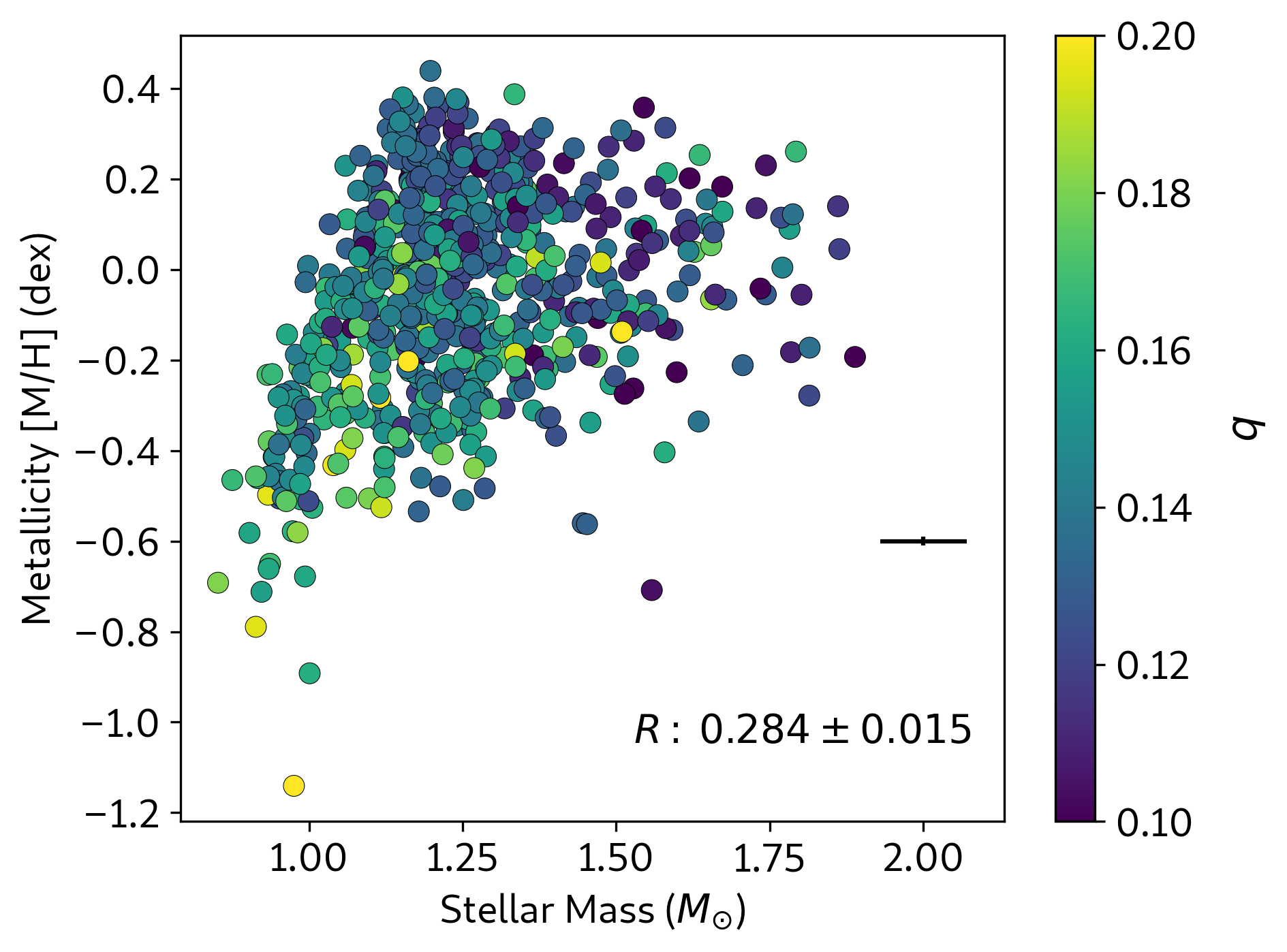}
    \caption{Stellar mass and bulk metallicity [M/H] for the 684 targets in our sample with chemical abundances from APOGEE DR17. Also shown is the Spearman coefficient ($R$) for between the two parameters, with 1 indicating a perfect, monotone correlation and 0 indicate the absence of any correlation.}
    \label{fig:fehmass_corr}
\end{figure}

Figure \ref{fig:fehmass_corr} shows the stellar mass and bulk metallicity of the 684 targets in our sample with chemical abundances from APOGEE DR17. As quantified by a Spearman rank correlation of $R=0.284\pm0.015$, a decrease in metallicity does not necessitate a decrease of stellar mass within our sample. Therefore, it is likely that there both mass and metallicity have separate influences upon the $\Delta\nu-q$ sequence as discussed in Section \ref{section:q}. 

\section{Identifying Incorrect Fits from the Triplet Rotational Splitting Assumption}\label{appendix:triplet}
To correct for incorrect fits from the three-component (triplet) assumption in Section \ref{section:triplet_assump}, we determine the contribution of each rotational multiplet from an existing BayesOpt solution of $(\Delta\Pi_1, q, \epsilon_g and \delta\nu_\mathrm{rot})$. This is done by removing synthetic modes of one particular rotational component at a time and recalculating the corresponding increase in the BayesOpt loss. The fractional contribution of different rotational components to the BayesOpt loss informs us which synthetic modes have been assigned.

\begin{figure}
    \centering
    \includegraphics[width=0.75\columnwidth]{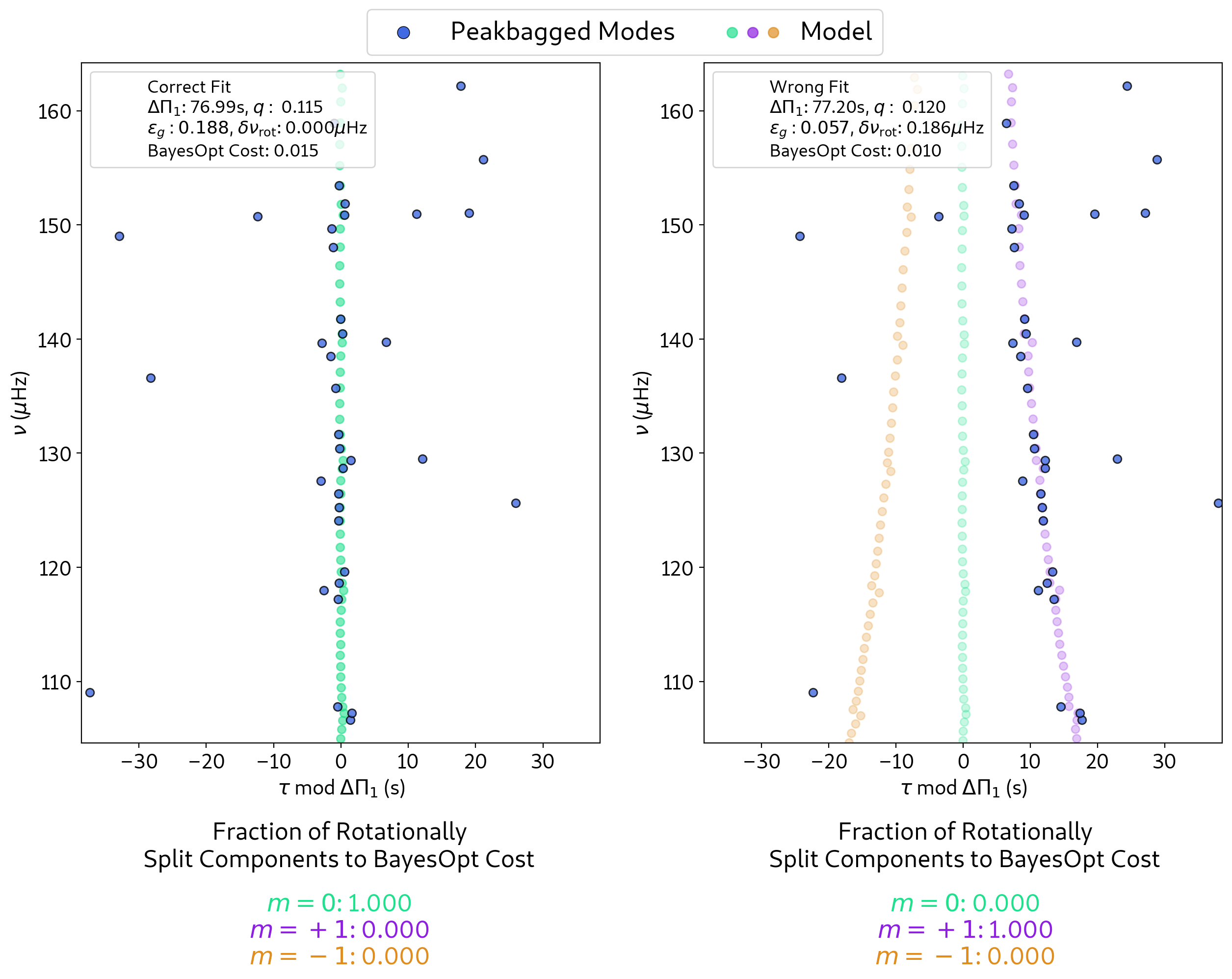}
    \caption{(\textit{Left}) The correct fit of synthetic mixed-mode frequencies to the peakbagged modes of KIC 4912198, a red giant viewed pole-on. \textit{(Right)} A solution with a comparable BayesOpt loss but incorrectly assigns the peakbagged modes to the $m=+1$ component. This inaccurate assignment is reflected in  the fractional contribution of each rotational component to the BayesOpt loss listed below each panel.}
    \label{fig:appendix_singlet}
\end{figure}

\begin{figure}
    \centering
    \includegraphics[width=0.75\columnwidth]{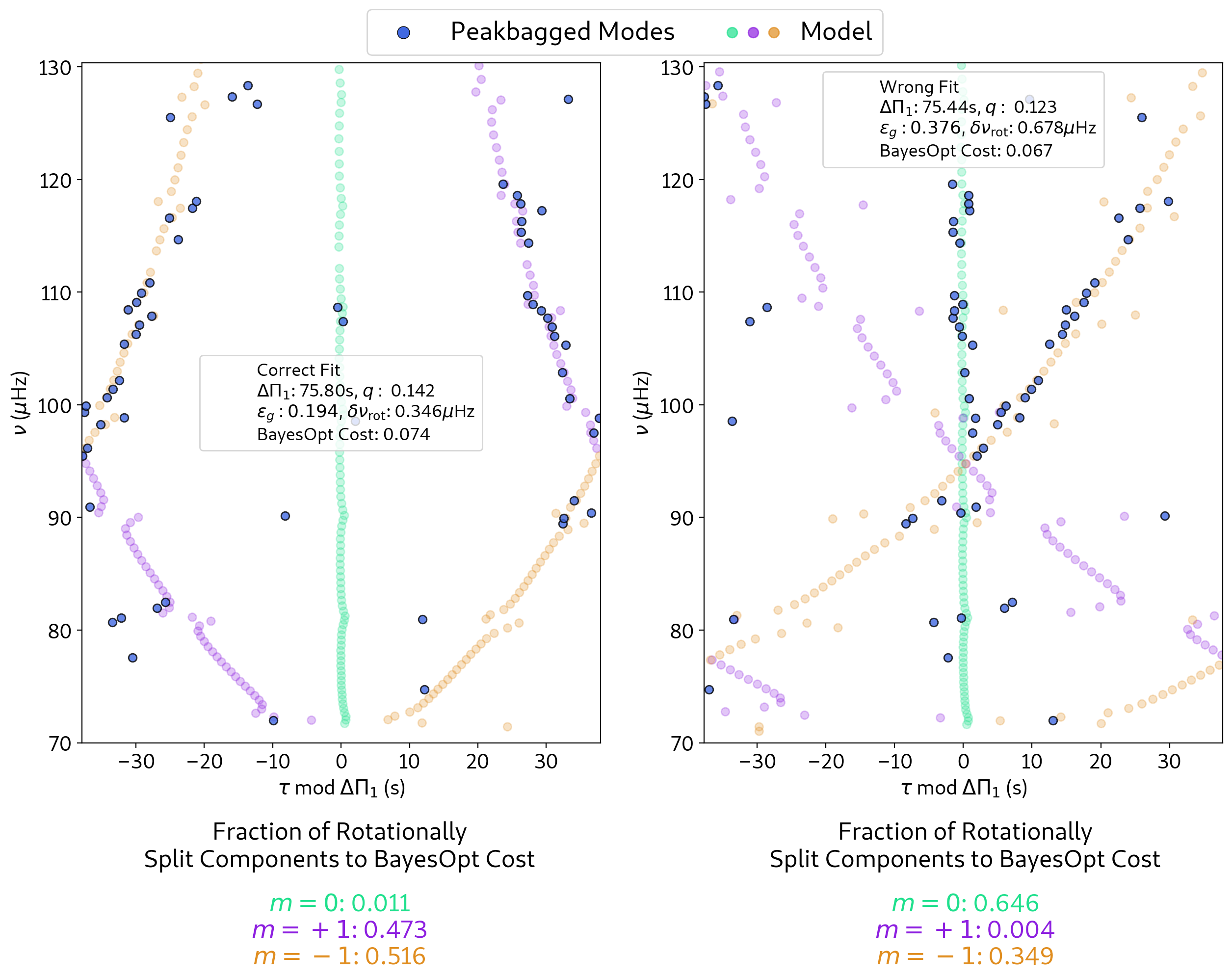}
    \caption{(\textit{Left}) The correct fit of synthetic mixed-mode frequencies to the peakbagged modes of KIC 8429977, a red giant viewed edge-on. \textit{(Right)} A solution with a comparable BayesOpt loss but incorrectly assigns the peakbagged modes to the $m=-1, 0$ components. Similar to Figure \ref{fig:appendix_singlet} this inaccurate assignment is reflected in the fractional contribution of each rotational component to the BayesOpt loss listed below each panel.}
    \label{fig:appendix_doublet}
\end{figure}

Figure \ref{fig:appendix_singlet} shows an example of the star KIC 4912198 that is viewed pole-on. The correct fit should have all contribution to the BayesOpt within the $m=0$ component. An incorrect assignment in the right panel of the Figure, however leads to all the loss contribution within the $m=+1$ component instead.
Figure \ref{fig:appendix_doublet} shows an example for the star KIC 8429977 that is viewed edge-on. The correct fit should have equal contribution to the BayesOpt loss within the $m=\pm1$ components, but an incorrect assignment leads to the $m=-1, 0$ components carrying most of the loss contributions instead. 

\begin{figure}
    \centering
    \includegraphics[width=0.75\columnwidth]{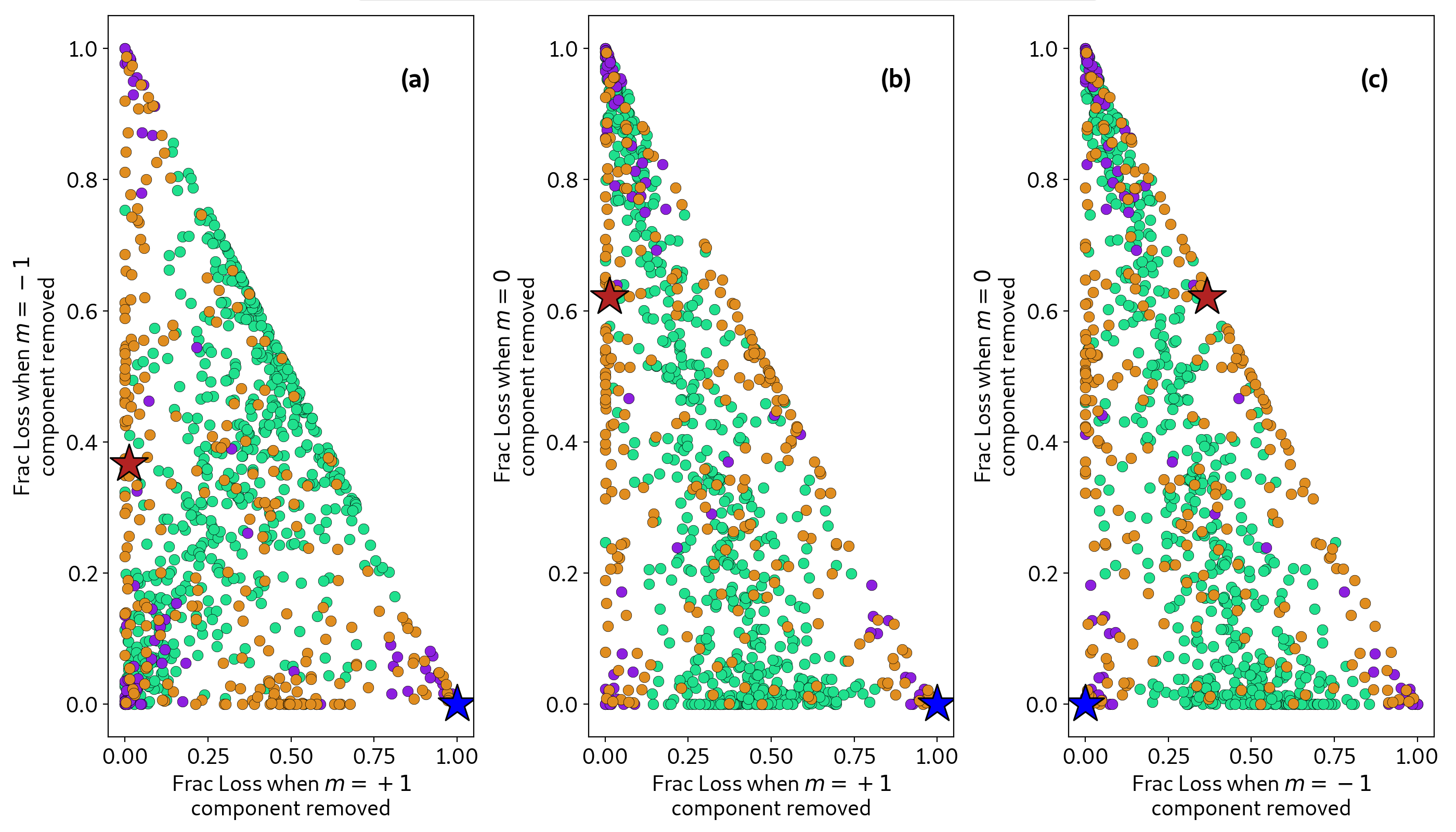}
    \caption{ Relative contributions of different rotational components to the BayesOpt loss. The sample in this plot has not been vetted, meaning that they are obtained directly from an automated run of BayesOpt over the sample in this work using the triplet rotational splitting assumption. The labels are provided by manual inspection of the fits. The incorrect fits from Figures \ref{fig:appendix_singlet} and Figures \ref{fig:appendix_doublet} are included. }
    \label{fig:appendix_triangle_raw}
\end{figure}

\begin{figure}
    \centering
    \includegraphics[width=0.75\columnwidth]{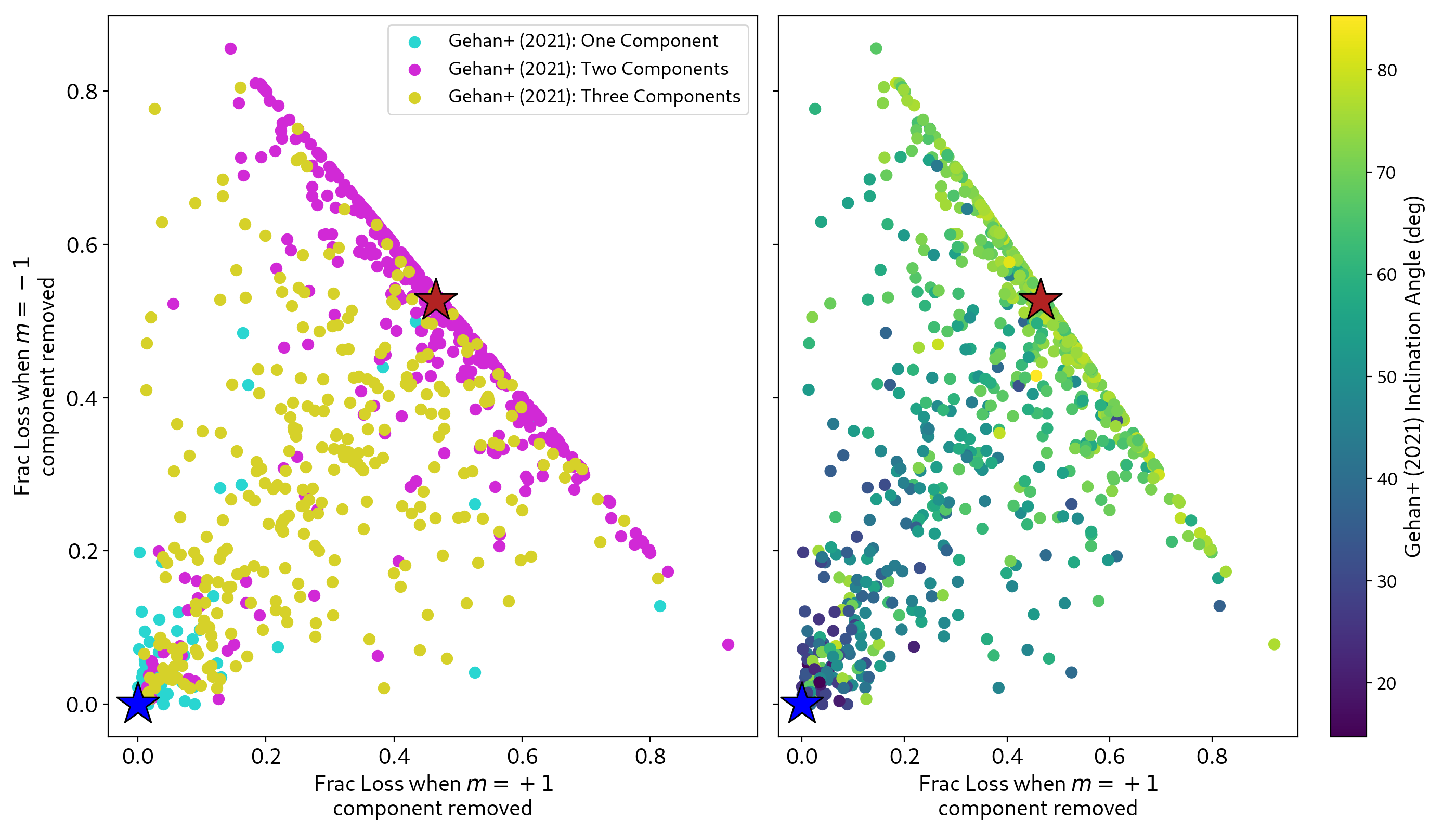}
    \caption{Similar to Figure \ref{fig:appendix_triangle_raw}a, but for the final results we report in our work. Here the incorrect fits have been rectified, leaving behind the correct fits that occupy a specific regime within this parameters space. The incorrect fits from Figures \ref{fig:appendix_singlet} and Figures \ref{fig:appendix_doublet} are included. To confirm the validity of our analysis, we color-coded points in the plots according to the number of rotationally-split components (\textit{left}) and stellar inclination angle as reported by \citet{Gehan_2018}.}
    \label{fig:appendix_triangle_final}
\end{figure}

To show this, we plot the relative contributions over the un-vetted results of our sample in Figure \ref{fig:appendix_triangle_raw}, where both incorrect and correct fits have been labelled by manual inspection. In contrast, we also show the equivalent of Figure \ref{fig:appendix_triangle_raw}a for our final, vetted sample in Figure \ref{fig:appendix_triangle_final} From these plots, it is evident that the regime occupied for correct and incorrect fits within this parameter space differs from one another. Therefore, any BayesOpt solution that produces a solution belonging outside the regime of good fits within this parameter space is subject to further inspection and re-optimized with the correct number of rotational components if necessary.

\bibliography{sample631}{}
\bibliographystyle{aasjournal}

\end{document}